\documentclass[fleqn,usenatbib]{mnras}

\usepackage{newtxtext,newtxmath}

\usepackage[T1]{fontenc}

\DeclareRobustCommand{\VAN}[3]{#2}
\let\VANthebibliography\thebibliography
\def\thebibliography{\DeclareRobustCommand{\VAN}[3]{##3}\VANthebibliography}

\usepackage{graphicx}	
\usepackage{amsmath}	
\usepackage{longtable}
\usepackage{supertabular}
\usepackage{pdflscape}
\usepackage{fix-cm}

\defcitealias{pfeifle2022bass}{P22}
\defcitealias{Asmus_diagnostic}{A15}

\title[An X-ray-MIR diagnostic for absorption]{Extending an X-ray {$\mathrm{\&}$} Mid-Infrared diagnostic for absorption in Active Galactic Nuclei to higher redshifts using \textit{NuSTAR} and \textit{XMM-Newton}}

\author[R. J. Turney et al.]{
Ryan J. Turney,$^{1}$\thanks{E-mail: ryan.turney@dunelm.org.uk}
C. L. Greenwell,$^{1}$
D. M. Alexander$^{1}$
\\
$^{1}$Centre for Extragalactic Astronomy, Department of Physics, Durham University, Durham, DH1 3LE, UK
}

\date{Accepted XXX. Received YYY; in original form ZZZ}

\pubyear{\the\year{}}

\begin{document}
\label{firstpage}
\pagerange{\pageref{firstpage}--\pageref{lastpage}}
\maketitle

\begin{abstract}
Determining the Compton thick fraction ($f_{CT}$) within the population of Active Galactic Nuclei (AGN) provides insight into the evolution and growth of Supermassive Black Holes. Measuring $f_{CT}$ requires methods of modelling the AGN column density ($N_{\mathrm{H}}$), typically through spectral fitting which requires high source counts, or estimation methods such as the band ratio. We extend an $N_{\mathrm{H}}$ estimation method using the ratio of observed X-ray and MIR luminosity ($\log\left({L}_{2-10\mathrm{keV}} / {L}_{12\mathrm{\mu m}}\right)$), previously applied in the local universe ($z<0.1$), to a higher redshift. We used a sample of 47 bright AGN from the \textit{NuSTAR} 80-month catalogue (NSS80) with a median redshift of $\langle z \rangle = 0.35$ as the high redshift comparison, and we applied X-ray spectral fitting to the \textit{NuSTAR} sources with physically motivated AGN models to provide $N_\mathrm{H}$ constraints. The high-$z$ relationship was found to be consistent with the same model as the low-$z$ comparison, but with modestly different parameters which we determined to be due to a higher X-ray luminosity within the NSS80 sample. We applied the model to the full NSS80 catalogue to determine a statistical measure of the observed Compton thick fraction at $f_{CT}=4.9^{+2.0}_{-2.3}\%$ to $1\sigma$ uncertainty, consistent with other recent low-$z$ \textit{Swift/BAT} measurements.

\end{abstract}




\section{Introduction}

Super-Massive Black Holes (SMBHs) are known to be at the centre of most galaxies. These SMBHs are thought to have primarily grown through phases of active accretion of dust, gas and stellar material \citep{ALEXANDER2025DecadeReview}; in this phase it is called an Active Galactic Nucleus \citep[AGN;][]{AGN..In..Name}.

AGN emit across the entire electromagnetic spectrum, with the accretion disk as the primary power source of the AGN \citep{ALEXANDER2025DecadeReview}. X-ray emission is a result of optical-UV photons inverse Compton scattered in what is commonly refered to as the corona \citep{opticalUV2024bass}. The emission of the corona follows a power-law up to the photoelectric cut-off at approximately 200 keV \citep{2018NuSTAR_HardBand} with a wide spread between samples \citep{photoelectric_cutoff}, before a sharp decline in differential photon flux following $N(E)\propto E^{-\Gamma}\exp{\left(-E/E_{cut}\right)}$ \citep{photoelectric_cutoff,corona_temp}. Additional soft X-ray emission below 1 keV is thought to arise from optical-UV photons up scattered off a warm corona close to the accretion disk which form the soft X-ray excess \citep{reflection,Petrucci_soft_2018}. Further X-ray emission originates from reflection, scattering and photoelectric absorption from the accretion disk or inner regions of the \lq dusty\rq$ $ torus \citep{AGN..In..Name}. The reprocessed and reflected X-ray emission forms a Compton hump which peaks at 20-30 keV \citep{revisited_unified}, in addition to a significant Iron K$\alpha$ emission line at 6.4 keV \citep{reflection,2018obscured}. Accretion disk photons are further absorbed and reprocessed by dust and gas within the \lq dusty\rq$ $ torus and polar outflows, before the photons are re-emitted in Mid-Infrared (MIR) energies \citep{AGNobscurationReview,Hoenig_2019}.

Along line of sight, the \lq dusty\rq$ $ torus attenuates AGN emission from the central regions due to a variety of absorption and scattering processes \citep{2018obscured}. X-rays are advantageous in observing AGN due to the ability to largely penetrate through the otherwise absorbing media, with hard X-rays ($\geq$10 keV) able to pass through the torus up to Compton thick (CT) obscuration \citep{AGN..In..Name}, defined at $\log(N_\mathrm{H}$/cm$^{-2})\gtrsim 24$\footnote{The CT boundary is strictly defined as $N_\mathrm{H} > 1.5 \times 10^{24}$ cm$^{-2}$, however in this work the approximation is used to follow \citet{pfeifle2022bass} and \citet{2018NuSTAR_HardBand}.}, where $N_\mathrm{H}$ is the line of sight column density. The heavy attenuation of soft X-rays in CT AGN from Compton scattering  \citep{scattering2004compton,2018obscured} can cause a $\sim2$ orders of magnitude decrease in observed soft X-ray emission \citep{thick2025compton}. This strong suppression of photons often results in a more pronounced Iron K$\alpha$ emission line within the spectra compared to that of an unobscured ($\log(N_\mathrm{H}$/cm$^{-2})\lesssim 22$) or Compton thin AGN ($23<\log(N_\mathrm{H}$/cm$^{-2})< 24$) \citep{BrandtW_xray_AGN}.

The MIR emission can be an effective indicator of intrinsic AGN output, due to the low optical depth of the torus and polar outflows at MIR wavelengths, resulting in little attenuation \citep{2018obscured}. The optical depth for MIR emission is wavelength dependant, resulting in increased attenuation for shorter wavelength emission along the line-of-sight. The polar outflows are thought to extend from the innermost regions of the torus to the SMBH, and can span distances of $10^2$ pc \citep{Hoenig_2019}. The torus emits across a wide range in dust temperatures of $T\sim10^2-10^3 $ K \citep{Hoenig_2019}, corresponding to thermal MIR emission from the dust at wavelengths of 3-30 $\mathrm{\mu m}$, with a significant amount of the MIR emission also originating from the polar outflows \citep{2018obscured,Hoenig_2019}.

AGN emission is often contaminated by the host galaxy, which can prove to be challenging when identifying low-luminosity AGN or AGN that reside in luminous galaxies \citep{2018obscured}. X-rays are important for distinguishing the AGN against the host galaxy, as there is relatively little contamination from astrophysical processes at X-ray energies within the galaxy \citep{AGN..In..Name,2018obscured}. MIR emission however can be heavily diluted from obscuring dust, heated by star-formation within the galaxy to temperatures on the order of $T=10^1-10^2\mathrm{K}$ \citep{Netzer_book,2018obscured}

Identifying CT AGN is crucial to develop a more complete census. The observed CT fraction is significantly less than what is predicted by Cosmic X-ray Background (CXB) estimates \citep{Gilli_2006,Treister_2009,Ricci_2015,Ananna_2019}, implying that a significant proportion of the AGN population is undetected or unidentified within X-ray AGN surveys. A more complete census would provide greater insight into the properties of AGN, including their accretion rates and evolution with the host galaxy \citep{2018obscured,Greenwell_Thesis,thick2025compton}. 

The coronal emission of X-rays and the reprocessed polar outflow and torus emission of MIR wavelengths both have a common origin from the optical-UV accretion disk photons. Such a property implies an intrinsic relation between the two bands. The work of \citet{Horst_2008} and \citet{Gandhi_2009} used the theoretical relation between the intrinsic X-ray and MIR emission to determine the measured correlation between the luminosities and infer properties about the torus models. Heavily obscured AGN have significantly lower observed $L_X/L_{MIR}$ than unobscured AGN, an effect that disappears when the X-ray luminosity is corrected for the measured X-ray absorption \citep{Horst_2008,Gandhi_2009}. For an obscured AGN, the X-rays are significantly attenuated by the torus, where the MIR emission is not, due to the low optical depth of the torus at MIR wavelengths. Further research from \citet{Asmus_diagnostic} and \citet{pfeifle2022bass} (here-on \citetalias{Asmus_diagnostic} and \citetalias{pfeifle2022bass} respectively) determined an estimation method using the ratio of observed 2-10\,keV X-ray and 12\,$\mathrm{\mu m}$ MIR luminosity to predict the column density of a source.
The prediction method (here-on the X-ray-MIR diagnostic) allows for sources to be categorised by their obscuration, even for a very limited number of counts, which can be used to identify CT sources that suffer from heavy X-ray attenuation \citepalias{pfeifle2022bass}. A majority of research into X-ray and MIR sources have focussed primarily on the local, low-$z$ universe. Observing AGN over a range of redshifts improves our understanding of AGN evolution, while also covering a broader range of luminosities \citep{Barger_2005}. Previous research from \citet{Laloux_2024_accretion} identified increasing average accretion rates with redshift, up to $z\sim3$, due to more available gas on average in a younger galaxy. Similar findings from \citet{Birchall_2022} even exhibit a difference in accretion rate and luminosity between AGN at $0.00<z<0.05$ and $0.25<z<0.35$, with higher accretion rates in the earlier universe. Probing AGN in the earlier universe and developing a more complete census will enable further understanding of accretion physics and AGN feedback with the host galaxy.

To advance the X-ray-MIR diagnostic, we use a higher $z$ sample of AGN to determine the validity of the X-ray-MIR diagnostic in the earlier universe. The X-ray-MIR diagnostic model will allow the column density to be predicted for sources with too few counts for spectral fitting, and provide an additional metric to identify CT sources instead of the band ratio. The X-ray-MIR diagnostic can then be applied to future catalogues with X-ray and MIR source data at higher redshifts, to determine a new measurement of the observed CT fraction ($f_{CT}$), and predict the distribution of AGN obscuration within the respective catalogue. We will primarily use the \citetalias{pfeifle2022bass} sample for the comparison at low $z$ in this paper due to the use of observed 12\,$\mathrm{\mu m}$ luminosity, and a similar selection method of X-ray selected AGN. The \citetalias{Asmus_diagnostic} sample is included for their alternative diagnostic method (see section~\ref{sec: linear diagnostic}), and the provided nuclear 12\,$\mathrm{\mu m}$ luminosity, which allows us to discuss the impact of the MIR AGN luminosity on the X-ray-MIR diagnostic fit (see section~\ref{sec: The X-ray-MIR diagnostic}).

The structure of the paper is as follows: Section~\ref{sec: Source sampling and catalogue description} describes the low-$z$ comparison samples of \citetalias{pfeifle2022bass} and \citetalias{Asmus_diagnostic} used in the paper, in addition to the selection methods used to produce the high redshift sample and the processing of the sources. Section~\ref{sec: Spectral fitting methods and calibration} details the modelling and spectral fitting of the sample. From here, the reduced sample is K-corrected in the MIR and fit to the X-ray-MIR diagnostic in section~\ref{sec: The X-ray-MIR diagnostic}. We analyse the results of fitting the sample to the X-ray-MIR diagnostic against the work of \citetalias{Asmus_diagnostic} and \citetalias{pfeifle2022bass} in Section~\ref{sec: mir results}, before applying the X-ray-MIR diagnostic to the NSS80 catalogue in Section~\ref{sec: Application to the NSS80}. Finally we determine the observed CT fraction distribution predicted by the X-ray-MIR diagnostic in Section~\ref{sec: CT fraction analysis} before we provide concluding remarks in Section~\ref{sec: Conclusions}. 

We assume a flat universe with cosmological parameters of H$_0=70$km s$^{-1}$ Mpc$^{-1}$ and $\Omega_{\Lambda 0}=0.73$ following the work of \citet{2018NuSTAR_HardBand}. All uncertainties are quoted to the $\pm1\sigma$ confidence level.

\section{Source sampling and catalogue description}
\label{sec: Source sampling and catalogue description}

Here we describe the high-$z$ \textit{NuSTAR} sample used within this paper in section~\ref{sec: Sample catalogue NSS80 description} before discussing the two low-$z$ comparison samples in section~\ref{sec: sample selection criteria}. Finally, we detail the data processing of the \textit{NuSTAR} and \textit{XMM-Newton} observations from the sample in section~\ref{sec: Data Processing }.

\subsection{The Sample Catalogues}
\label{sec: Sample catalogue NSS80 description}

In this paper we use a higher redshift sample selected from the \textit{NuSTAR} 80-month serendipitous survey (here-on the NSS80). The NSS80 covers 894 unique fields and has detected 1274 hard X-ray sources, and spans $\sim$36 deg$^2$ with a median redshift of $\langle z \rangle = 0.56$ \citep{2024ApJS..273...20G}. While \textit{NuSTAR} covers the energy range of 3-79keV, the primary focus of the NSS80 catalogue is within the 3-24keV band, due to the optimal sensitivity of the \textit{NuSTAR} telescope \citep{Harrison_2013} within that range. The NSS80 builds upon the work of the 40-month catalogue \citep[the NSS40;][]{Lansbury_2017_nss40} with an an additional 822 new sources.

We further utilised the \textit{XMM-Newton} 4XMM-DR10 catalogue \citep{xmm_catalogue} to extend the spectral coverage of the NSS80 sources due to its sensitivity in the 0.5-12 keV energy range across the PN and MOS cameras. For the additional Infrared (IR) data, we used the the Wide-field Infrared Survey Explorer (\textit{WISE}) \citep{Wright_2010}. \textit{WISE} uses four observational bands W1-W4, corresponding to effective wavelengths of 3.4,4.6,12 and 22\,$\mathrm{\mu m}$ respectively \citep{Marocco_2021}. 

\subsection{Comparison sample selection}
\label{sec: Comparison sample selection}

The work of \citetalias{Asmus_diagnostic} used a sample of 152 local MIR and X-ray selected AGN. \citetalias{Asmus_diagnostic} builds off the subarcsecond-resolution MIR atlas of AGN \citep{atlas_asmsus}, containing X-ray sources from the 54 and 70 month \textit{Swift/BAT} catalogue \citep{Cusumano_2010_54month,Baumgartner_2013}. The MIR counterpart data was taken from several ground based instruments including VISIR \citep{lagage2004_visir}, Gemini/Michelle \citep{Glasse_1997} and COMICS \citep{Kataza_2000}. Sources with inconsistent flux measurements, or poor quality spectra were removed, leaving the final 152. These sources were used to determine a linear estimation model between $2$-$10\mathrm{keV}$ X-ray and 12\,$\mathrm{\mu m}$ MIR AGN flux, for obscured sources with $\log(N_\mathrm{H}$/cm$^{-2})> 22.8$.

The X-ray-MIR diagnostic for absorption developed by \citetalias{pfeifle2022bass} built upon the work of \citetalias{Asmus_diagnostic} with a sample of 456 local AGN. The sample was developed from the \textit{Swift/BAT} 70-month catalogue \citep{Baumgartner_2013} with measurements of $N_{\mathrm{H}}$ and \textit{WISE} MIR luminosity from \citet{BAT_V_XRAY_2017} and \citet{Ichikawa_2017} respectively. The \citetalias{pfeifle2022bass} sample further applied an upper redshift limit of $z<0.1$ to account for redshift effects in the \textit{WISE} MIR magnitudes \citepalias{pfeifle2022bass}. The work of \citetalias{pfeifle2022bass} used a power-law model to model the luminosity ratio ($\log\left( {L}_{2-10\mathrm{keV}} / {L}_{12\mathrm{\mu m}}\right)$) to estimate the column density of local AGN sources.

\subsection{Sample selection}
\label{sec: sample selection criteria}

To test the $L_X/L_{MIR}$ ratio as a diagnostic for the identification of heavily obscured and CT AGN, we required source measurements of both $N_{\mathrm{H}}$ and $L_{2-10\mathrm{keV}}$ through spectral fitting. Our primary sample within this paper is the NSS80 from \citet{2024ApJS..273...20G}. In order to model the spectra of the sources, we applied selection criteria which largely follow the work of \citet{2018NuSTAR_HardBand}. This paper primarily focussed on the identification and spectral fitting of extreme CT sources with \textit{NuSTAR} and soft X-ray telescope counterparts to be used as a comparison to the CT fraction predicted by the Cosmic X-ray Background (CXB).

For the X-ray spectral fitting, we select only bright sources with a lower bound cut on the \textit{NuSTAR} net deblended hard-band flux $F_{8-24\mathrm{keV}} >7\times 10^{-14} $\,erg\,s$^{-1}$\,cm$^{-2}$. The galactic latitude was also limited to $|b| > 10 ^\circ$ to reduce Galactic obscuration and confusion of the X-ray sources. All blazars and Galactic sources were also removed from the candidate sample. The minimum full-band (3-24keV) deblended net counts of \textit{NuSTAR} was set at $N\geq100$ to ensure there were sufficiently high counts for the \textit{NuSTAR} spectral fitting. As discussed in in section~\ref{sec: Sample catalogue NSS80 description}, we selected \textit{XMM-Newton} as the soft X-ray counterpart to the \textit{NuSTAR} sources. NSS80 sources with \textit{XMM-Newton} as the adopted X-ray coverage of the NSS80 were selected for the candidate sample and matched to the 4XMM-DR10 catalogue \citep{xmm_catalogue}. If multiple \textit{XMM-Newton} observations were available, the observation with the closest time stamp to \textit{NuSTAR} and an exposure time of $>10\mathrm{ks}$ was used.

The NSS80 includes \textit{WISE} counterpart MIR magnitudes from statistically cross-matching sources using NWAY \citep{nway_2018}. The \textit{WISE} counterpart MIR magnitudes provided in the NSS80 catalogue \citep{2024ApJS..273...20G} were required to determine the X-ray to MIR ratio ($\log\left( {L}_{2-10\mathrm{keV}} / {L}_{12\mathrm{\mu m}}\right)$) used for the X-ray-MIR diagnostic. We kept candidate sources with W3 and associated uncertainty, in addition to either W4 with uncertainty, or W2 with uncertainty. The W4 or W2 detection requirement is needed to apply an approximate K-correction to the W3 flux (see section~\ref{sec:K-correction}). The final constraints required an optical counterpart with good quality spectrum to ensure there was a valid redshift measurement for the later spectral fitting.

From the application of these cuts, the final sample used in the X-ray spectral fitting comprises 48 sources. Following spectral analysis in section~\ref{sec: Source checks} (and Appendix~\ref{sec: Source Variability}), we remove one additional source due to evidence of extreme variability, reducing the sample to 47 sources. This sample will be referred to as the NSS80 sample from now on; details of the NSS80 sample source observations can be seen in Appendix \ref{Appendix: Source info}.

\begin{figure}
	\includegraphics[width=\columnwidth]{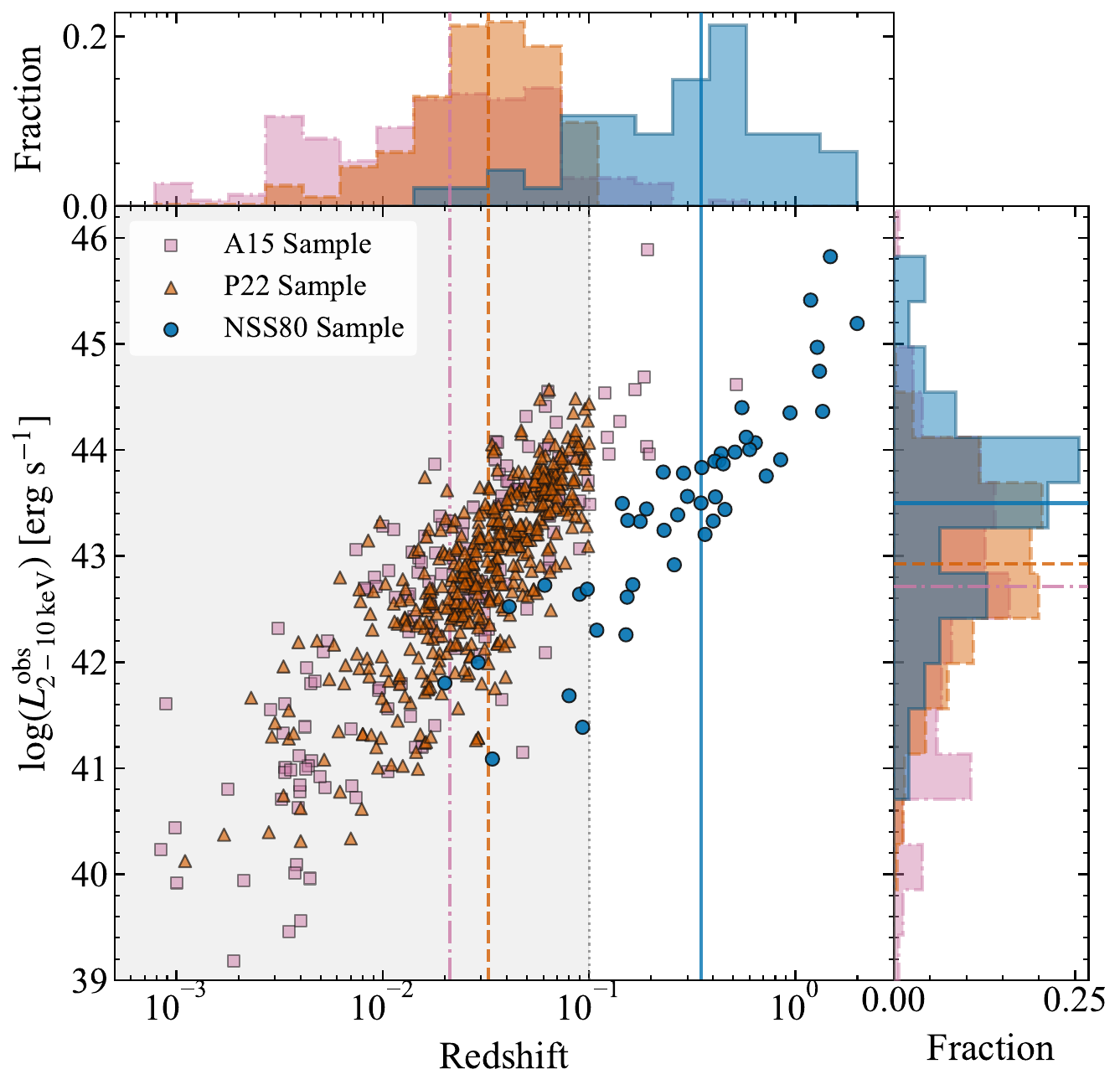}
    \caption{The distribution of  2-10\,keV luminosity ($L_{2-10\mathrm{keV}}$) against redshift for the NSS80 sample of 47 sources (blue circles), 456 \citetalias{pfeifle2022bass} sources (orange triangles) and 151 \citetalias{Asmus_diagnostic} sources (pink squares). The top panel plots the redshift distribution for the NSS80 (blue), \citetalias{pfeifle2022bass} (orange dashed) and \citetalias{Asmus_diagnostic} (pink dot-dashed), with the right panel showing the luminosity distribution. The median redshift for each sample can be seen with the vertical lines, with the median 2-10\,keV luminosity shown as horizontal lines on the right panel (dash-dot for \citetalias{Asmus_diagnostic}, dashed for \citetalias{pfeifle2022bass} and solid for the NSS80). The upper redshift limit imposed within \citetalias{pfeifle2022bass} can be seen in the grey region at $z<0.1$. The NSS80 sample is at a significantly higher redshift and median X-ray luminosity than the other two samples.}
    \label{fig:sample redshift graphs}
\end{figure}

Fig.~\ref{fig:sample redshift graphs} plots the observed 2-10\,keV X-ray luminosity against redshift distribution for the three samples discussed within this paper. The NSS80 sample probes higher-redshift sources with a median of $\langle z_{\mathrm{Nu}} \rangle = 0.35$. The \textit{Swift/BAT} samples of \citetalias{Asmus_diagnostic} and \citetalias{pfeifle2022bass} show a significantly lower redshift for AGN in the local universe, with $\langle z_{\mathrm{A15}} \rangle =0.02$ and $\langle z_{\mathrm{P22}} \rangle = 0.03$. The NSS80 therefore provides a sufficient comparison at high $z$ when fitting and extending the X-ray-MIR diagnostic.

\subsection{Data processing}
\label{sec: Data Processing } 
We follow standard analysis threads with nustardas v2.1.5 for the processing of \textit{NuSTAR} with the calibration database CALDB version 20251215. The processing of all \textit{XMM-Newton} sources uses xmmsas v22.1.0.

The aperture size for the extraction of the X-ray data used in the spectral fitting was taken at 30-45" for \textit{NuSTAR}, with the 30" aperture for objects close to another source or which were located the edge of the image. For \textit{XMM-Newton}, an aperture size of 12-15" was used with similar criteria to \textit{NuSTAR} following \citet{extreme2017nustar}. 
The background aperture for \textit{NuSTAR} and \textit{XMM-Newton} were selected to be larger than the source aperture, within the same chip as the source and have no sources visible within the aperture itself.

Following the similar \textit{NuSTAR}-\textit{XMM-Newton} spectral fitting approach of \citet{2018NuSTAR_HardBand}, the spectral counts were binned at a minimum of 5 counts per bin for both \textit{XMM-Newton} and \textit{NuSTAR}. 

\section{X-ray Spectral fitting methodology}
\label{sec: Spectral fitting methods and calibration}

In this section we describe the physically motivated AGN models used for X-ray spectral fitting the NSS80 sources which are used to constrain the $N_{\mathrm{H}}$ and $L_{2-10\mathrm{keV}}$ (section~\ref{sec: Spectral Modelling}). We then check the fit sources for potential variability between observations in section~\ref{sec: Source checks} to reduce any effect on the X-ray-MIR diagnostic model fitting in section~\ref{sec: The X-ray-MIR diagnostic}. Finally we discuss the spectral fit NSS80 sample in section~\ref{sec: sample discussion}. 

\subsection{X-ray spectral fitting}
\label{sec: Spectral Modelling}

This work closely follows the approach of \citet{Greenwell_Thesis} in spectral fitting models, which used selection techniques and analysis methods to identify \lq missing\rq\ or rare AGN, including CT sources. Two models were used to broadly represent unobscured and obscured sources. Due to the low number of source counts within the sample, the best fit model of most sources (see section~\ref{sec: Source checks}) was determined using Cash statistics \citep[cstat;][]{cash_1979}.

The initial model adopts a simple power-law fit assuming emission from the corona, with additional absorption (\texttt{Tbabs}; updated photoionisation cross sections and the processing of interstellar dust grains; \citealp{Wilms_2000_tbabs}) and scattering (\texttt{cabs}; models the optically-thin Compton scattering out of the line-of-sight) from the Milky-Way. The column density of the MW was taken from the N$\mathrm{_H}$ HEAsoft tool\footnote{\url{https://heasarc.gsfc.nasa.gov/cgi-bin/Tools/w3nh/w3nh.pl}}, and fixed when spectral fitting. The model includes further photoelectric absorption at the host galaxy and AGN down the line-of-sight (\texttt{zwabs}), and a cross-instrument calibration constant fixed at 1 and 0.93 for \textit{NuSTAR} and \textit{XMM-Newton} respectively, using the values from \citet{extreme2017nustar}. The form of the model in XSPEC terminology is as follows: 
\vspace{2mm}

\centerline{\texttt{constant$\times$Tbabs$\times$cabs$\times$zwabs$\times$powerlaw}} 

\vspace{2mm}
The secondary, more complex model is the linked-$N_H$ \texttt{Borus02} model (here-on Borus) \citep{2018Borus}. \texttt{Borus02} models the reprocessed emission from the torus which includes Compton scattering and fluorescent line emission, such as the Iron K$\alpha$ line, in addition to an intrinsic power-law from the corona. The covering factor ($\cos{\theta_\mathrm{tor}}$) was fixed at 0.5, the solar abundance fixed at 1 and the viewing angle ($\cos{\theta_\mathrm{inc}}$) at 0.05, corresponding to a line-of-sight through the torus of a typical obscured AGN \citep{Greenwell_Thesis}. 

The first component (\texttt{constant$_1\times$Tbabs$\times$zTbabs}) models the absorption from both the MW and the host galaxy of the AGN, with a cross-calibration constant identical to the power-law model. The host galaxy absorption was approximated to be the same as the MW for all sources. Similar to the power-law, we fixed the host galaxy and MW absorption when spectral fitting. (\texttt{zTbabs$\times$cabs$\times$cutoffpl}) represents the line-of-sight absorption from the host galaxy. The final component \texttt{(constant$_2 \times$cutoffpl)} accounts for any leaked scattering that is not absorbed by the torus. The linked model ties the line-of-sight column density to the average column density, with the \texttt{cutoffpl} normalisation tied to the normalisation of \texttt{Borus02}. The \texttt{constant$_2$} component is free to vary. The linked Borus model is limited to modelling a column density of $\log(N_\mathrm{H}$/cm$^{-2})\geq22.0$. The secondary model in XSPEC terminology is:
\vspace{2mm}

\centerline{\texttt{constant$_1\times$Tbabs$\times$zTbabs$\times$}}

\centerline{\texttt{(Borus02$+$zTbabs$\times$cabs$\times$cutoffpl$+$constant$_2 \times$cutoffpl)}}

\vspace{2mm}
To account for sources with varying X-ray flux between \textit{NuSTAR} and \textit{XMM-Newton} observations, the cross-calibration constant (\texttt{constant/constant$_{1}$}) was allowed to vary. Freeing the component was done only if the cstat fit significantly improved with a varying \texttt{constant$_{(1)}$}, and if the time difference between observations was on the order of years. If the source was found to be varying, the most recent observation was treated as the \lq current\rq$ $ AGN flux and had a fixed \texttt{constant$_{(1)}$}. The other telescope's calibration constant was then allowed to vary. The results from the spectral fitting can be seen in Table~\ref{tab:Results}.              

\subsection{Source analysis}
\label{sec: Source checks}

We applied further analysis to the sample of 48 sources to ensure the best fit of each source. the Borus model was found to often fit an unobscured source as Compton thin, by exaggerating the flux from the leaked component of the model. The \texttt{cflux} component was used to model the flux in leaked component to determine if the fit was unrealistic \citep{Winter_2009}, which was found to be the case for NuSTARJ021435-0042.7 and NuSTARJ121426+1403.1 with $> 50\%$ of the flux from the whole spectra. As such, the power-law model was selected as the correct fit for them both despite the slightly worse cstat value. 

We also examined the variability of the 48 NSS80 sources using the \textit{Swift-XRT} 2SXPS catalogue \citep{2SXPS_catalogue}, which identified NuSTARJ024144+0512.3 as a potentially highly variable source. We therefore elected to remove NuSTARJ024144+0512.3 from the sample. A more detailed reasoning of the source removal can be found in Appendix~\ref{sec: Source Variability}.

\onecolumn  
\renewcommand{\arraystretch}{1.5}
\setlength{\LTcapwidth}{\textwidth} 

\begin{longtable}{llccccc} 
\caption{Spectral fit results for the NSS80 sample.} \label{tab:Results} \\ 
\hline 
Source & Model & $\log(N_\mathrm{H}/\mathrm{cm}^{-2})$ & $\Gamma$ & $\log(L^\mathrm{obs}_{2-10\mathrm{keV}})$ & $\log(F^\mathrm{obs}_{2-10\mathrm{keV}})$ & Fit (C/n) \\ 
(1) & (2) & (3) & (4) & (5) & (6) & (7) \\ 
\hline 
\endfirsthead 

\caption[]{Spectral fit results for the NSS80 sample (Continued).} \\ 
\hline 

Source & Model & $\log(N_\mathrm{H}/\mathrm{cm}^{-2})$ & $\Gamma$ & $\log(L^\mathrm{obs}_{2-10\mathrm{keV}})$ & $\log(F^\mathrm{obs}_{2-10\mathrm{keV}})$ & Fit (C/n) \\ 
(1) & (2) & (3) & (4) & (5) & (6) & (7) \\ 
\hline 
\endhead 

    \hline
    \multicolumn{7}{r}{Continued} \\
    \endfoot

    \hline
    \multicolumn{7}{r}{%
         \begin{minipage}{\dimexpr\textwidth-2\tabcolsep\relax}
          \vspace{2mm}
          $^{a}$ Sources with a free \texttt{constant/constant$_1$} parameter.\\
          $^{b}$ No available \textit{WISE} W4 magnitude or error data available.\\
          $^{c}$ Source removed due to potential high variability between observations.\\
          \textbf{Notes:} (1) The \textit{NuSTAR} source name, as in NSS80. 
          (2) Description of whether the best fit model was Borus or Power-law (`Pow'). 
          (3) Modelled column density from the best fit with $\pm 1\sigma$ uncertainty. 
          (4) Photon index from the best model fit. Some sources were fixed to $\Gamma=1.9$ if it significantly improved the statistical fit, or if $\Gamma$ was fitting to extreme values ($\Gamma \lesssim 1.3$). These are indicated by (1.9).
          (5) Observed 2-10\,keV luminosity (in erg\,s$^{-1}$) calculated using the XSPEC Lumin function for the best-fit model. 
          (6) Observed 2-10\,keV flux (in units of erg\,s$^{-1}$\,cm$^{-2}$) calculated using the XSPEC Flux function for the best-fit model.
          (7) Fit statistic describing cstat/D.o.F of the best fit model.  
         \end{minipage}%
    } \\
    \endlastfoot

        NuSTARJ001443+8131.9$^{\mathrm{b}}$ &Pow & $21.79^{+0.19}_{-0.18}$ & $1.84^{+0.10}_{-0.10}$ & $43.20^{+0.08}_{-0.10}$ & $-12.72^{+0.13}_{-0.21}$ & $78/65$ \\
        NuSTARJ012618+2559.2 &Pow & $<20.85$ & $1.72^{+0.13}_{-0.06}$ & $43.56^{+0.02}_{-0.03}$ & $-13.18^{+0.02}_{-0.04}$ & $188/204$ \\
        NuSTARJ012633+2556.2$^{\mathrm{b}}$&Pow & $<20.2$ & $2.17^{+0.06}_{-0.06}$ & $43.90^{+0.01}_{-0.02}$ & $-12.95^{+0.02}_{-0.02}$ &  $290/271$ \\
        NuSTARJ021435-0042.7 &Pow & $<20.59$ & $2.09^{+0.10}_{-0.04}$ & $43.97^{+0.02}_{-0.03}$ & $-12.90^{+0.03}_{-0.04}$ & $134/131$ \\
        NuSTARJ022742+3331.5 &Pow & $21.65^{+0.04}_{-0.05}$ & $1.60^{+0.06}_{-0.06}$ & $42.64^{+0.01}_{-0.01}$ & $-12.65^{+0.02}_{-0.02}$ & $302/337$ \\
        NuSTARJ023229+2023.7 &Borus & $23.05^{+0.02}_{-0.04}$ & $1.90^{+0.13}_{-0.13}$ & $42.00^{+0.04}_{-0.09}$ &$-12.28^{+0.06}_{-0.10}$ &  $179/147$ \\
        NuSTARJ024144+0512.3$^{\mathrm{c}}$ &Borus & $23.68^{+0.03}_{-0.02}$ & $1.43^{+0.06}_{-0.02}$ & $43.01^{+0.01}_{-0.03}$ & $-12.01^{+0.01}_{-0.04}$ & $318/314$ \\
        NuSTARJ031015-7651.5 &Pow & $<20.86$ & $1.93^{+0.04}_{-0.04}$ & $45.41^{+0.01}_{-0.01}$ &$-12.46^{+0.01}_{-0.01}$ &  $287/308$ \\
        NuSTARJ032459-0256.2$^{\mathrm{a}}$ &Pow & $<20.28$ & $1.75^{+0.05}_{-0.05}$ & $41.80^{+0.02}_{-0.02}$ & $-12.74^{+0.03}_{-0.04}$ & $213/188$ \\
        NuSTARJ042613-5714.9 &Pow & $<20.45$ & $2.15^{+0.04}_{-0.03}$ & $42.69^{+0.01}_{-0.02}$ & $-12.71^{+0.02}_{-0.02}$ & $351/339$ \\
        NuSTARJ045348+0408.6 &Borus & $23.87^{+0.18}_{-0.20}$ & $(1.9)$ & $44.36^{+0.09}_{-0.17}$ & $-13.12^{+0.02}_{-0.14}$ & $119/124$ \\
        NuSTARJ051937-3233.3$^{\mathrm{b}}$&Pow & $<21.3$ & $1.7^{+0.15}_{-0.10}$ & $44.07^{+0.01}_{-0.03}$ & $-13.12^{+0.02}_{-0.06}$ & $186/203$ \\
        NuSTARJ052531-4557.8 &Pow & $<20.63$ & $1.76^{+0.02}_{-0.02}$ & $45.82^{+0.01}_{-0.01}$ & $-12.21^{+0.01}_{-0.01}$ & $460/428$ \\
        NuSTARJ090053+3856.3 &Borus & $23.40^{+0.05}_{-0.05}$ & $1.45^{+0.20}_{-0.02}$ & $43.79^{+0.01}_{-0.03}$ & $-12.27^{+0.01}_{-0.01}$ & $151/125$ \\
        NuSTARJ094910+0022.9 &Borus & $24.36^{+0.10}_{-0.20}$ & $(1.9)$ & $41.39^{+0.04}_{-0.15}$ & $-13.89^{+0.06}_{-0.15}$ & $243/208$ \\
        NuSTARJ102345+0044.1 &Borus & $23.49^{+0.10}_{-0.06}$ & $1.45^{+0.16}_{-0.16}$ & $43.56^{+0.03}_{-0.17}$ & $-12.70^{+0.10}_{-0.06}$ & $362/309$ \\
        NuSTARJ103135-4206.0 &Borus & $23.74^{+0.03}_{-0.02}$ & $1.45^{+0.07}_{-0.04}$ & $42.72^{+0.03}_{-0.06}$ & $-12.18^{+0.01}_{-0.02}$ & $469/427$ \\
        NuSTARJ103456+3939.6$^{\mathrm{a}}$ &Borus & $24.14^{+0.03}_{-0.02}$ & $2.55^{+>2.6}_{-0.06}$ & $42.26^{+0.07}_{-0.04}$ & $-13.37^{+0.06}_{-0.07}$ & $394/325$ \\
        NuSTARJ112745-2911.0$^{\mathrm{a}}$&Pow & $<20.5$ & $2.29^{+0.09}_{-0.08}$ & $43.44^{+0.03}_{-0.03}$ & $-12.59^{+0.03}_{-0.04}$ & $271/211$ \\
        NuSTARJ115658+5508.2 &Borus & $24.29^{+0.06}_{-0.06}$ & $2.59^{+>2.6}_{-0.51}$ & $41.68^{+0.02}_{-0.21}$ & $-13.39^{+0.02}_{-0.14}$ &  $79/64$ \\
        NuSTARJ115912+4232.6 &Pow & $22.77^{+0.05}_{-0.05}$ & $1.51^{+0.11}_{-0.11}$ & $43.33^{+0.01}_{-0.02}$ & $-12.54^{+0.01}_{-0.03}$ & $179/219$ \\
        NuSTARJ120348+4428.2$^{\mathrm{b}}$&Pow & $22.13^{+0.14}_{-0.20}$ & $1.65^{+0.07}_{-0.07}$ & $45.19^{+0.02}_{-0.02}$ & $-13.07^{+0.02}_{-0.03}$ & $278/263$ \\
        NuSTARJ121355+1404.4 &Borus & $22.42^{+0.07}_{-0.09}$ & $2.00^{+0.06}_{-0.07}$ & $43.34^{+0.01}_{-0.05}$ & $-12.47^{+0.01}_{-0.06}$ & $295/253$ \\
        NuSTARJ121426+1403.1 &Pow & $<20.63$ & $1.82^{+0.05}_{-0.02}$ & $44.97^{+0.01}_{-0.02}$ & $-12.96^{+0.01}_{-0.03}$ & $408/374$ \\
        NuSTARJ124946+2629.0 &Pow & $<20.41$ & $2.07^{+0.10}_{-0.06}$ & $43.89^{+0.02}_{-0.02}$ & $-12.89^{+0.02}_{-0.03}$ & $283/258$ \\
        NuSTARJ132231-1647.4 &Pow & $<20.49$ & $2.32^{+0.09}_{-0.05}$ & $43.50^{+0.02}_{-0.02}$ & $-13.16^{+0.02}_{-0.03}$ & $267/269$ \\
        NuSTARJ133401-3403.6$^{\mathrm{a}}$&Pow & $21.80^{+0.04}_{-0.04}$ & $1.77^{+0.12}_{-0.12}$ & $43.50^{+0.01}_{-0.03}$ & $-12.24^{+0.01}_{-0.03}$ & $363/344$ \\
        NuSTARJ134934-3025.5 &Borus & $22.53^{+0.19}_{-0.19}$ & $(1.9)$ & $42.73^{+0.05}_{-0.06}$ & $-13.10^{+0.04}_{-0.08}$ & $261/243$ \\
        
        NuSTARJ140536+2551.7 &Pow & $<21.05$ & $2.27^{+0.20}_{-0.10}$ & $44.35^{+0.03}_{-0.03}$ & $-13.39^{+0.03}_{-0.05}$ & $192/170$ \\
        NuSTARJ141809+2500.7$^{\mathrm{b}}$&Pow & $21.80^{+0.22}_{-0.24}$ & $(1.9)$ & $43.75^{+0.07}_{-0.06}$ & $-13.59^{+0.11}_{-0.02}$ & $100/83$ \\
        NuSTARJ141914+2104.2$^{\mathrm{b}}$&Borus & $22.97^{+0.13}_{-0.12}$ & $(1.9)$ & $42.30^{+0.03}_{-0.05}$ & $-13.15^{+0.03}_{-0.05}$ & $167/151$ \\
        NuSTARJ143026+4159.9 &Pow & $20.73^{+0.23}_{-0.49}$ & $1.94^{+0.06}_{-0.06}$ & $43.83^{+0.02}_{-0.02}$ & $-12.78^{+0.02}_{-0.02}$ & $226/235$ \\
        NuSTARJ150646+0346.2 &Borus & $24.38^{+0.04}_{-0.03}$ & $2.16^{+0.19}_{-0.11}$ & $41.09^{+0.01}_{-0.10}$ & $-13.30^{+0.01}_{-0.13}$ & $271/235$ \\
        
        NuSTARJ151653+5617.7 &Pow & $22.05^{+0.14}_{-0.21}$ & $(1.9)$ & $44.74^{+0.03}_{-0.04}$ & $-13.21^{+0.04}_{-0.04}$ & $207/141$ \\
        NuSTARJ152722+3600.2 &Borus & $23.06^{+0.06}_{-0.05}$ & $(1.9)$ & $43.39^{+0.02}_{-0.03}$ & $-12.88^{+0.03}_{-0.02}$ & $142/103$ \\
        NuSTARJ153640+5435.1 &Borus & $23.07^{+0.06}_{-0.04}$ & $(1.9)$ & $43.87^{+0.02}_{-0.02}$ & $-12.85^{+0.01}_{-0.02}$ & $173/164$ \\
        NuSTARJ161445+4659.3 &Borus & $23.71^{+0.05}_{-0.05}$ & $2.31^{+0.23}_{-0.22}$ & $42.61^{+0.01}_{-0.08}$ & $-13.11^{+0.01}_{-0.07}$ & $299/318$ \\
        NuSTARJ165105-0129.4 &Borus & $22.84^{+0.04}_{-0.04}$ & $1.63^{+0.08}_{-0.07}$ & $42.52^{+0.01}_{-0.2}$ & $-12.06^{+0.01}_{-0.02}$ & $309/326$ \\
        NuSTARJ183442+3237.9$^{\mathrm{a}}$&Pow & $21.57^{+0.21}_{-0.49}$ & $1.61^{+0.05}_{-0.17}$ & $43.98^{+0.04}_{-0.07}$ &$-12.95^{+0.05}_{-0.07}$ &  $132/105$ \\
        
        NuSTARJ192215-5844.9 &Pow & $<20.98$ & $1.89^{+0.19}_{-0.11}$ & $43.24^{+0.03}_{-0.04}$ & $-12.95^{+0.04}_{-0.04}$ & $203/203$ \\
        NuSTARJ193246-7233.8 &Borus & $22.11^{+0.04}_{-0.03}$ & $1.67^{+0.05}_{-0.06}$ & $43.78^{+0.02}_{-0.02}$ & $-12.59^{+0.02}_{-0.01}$ & $328/287$ \\
        NuSTARJ194234-1011.9$^{\mathrm{a,b}}$&Borus & $24.32^{+0.22}_{-0.27}$ & $1.9^{+0.15}_{-0.24}$ & $43.91^{+0.08}_{-0.04}$ & $-13.53^{+0.05}_{-0.13}$ & $293/204$ \\
        NuSTARJ200755-4439.7 &Borus & $23.84^{+0.11}_{-0.09}$ & $1.75^{+0.31}_{-0.25}$ & $43.44^{+0.01}_{-0.21}$ & $-13.20^{+0.01}_{-0.19}$ & $154/146$ \\
        NuSTARJ202226+6141.3$^{\mathrm{a,b}}$&Pow & $21.56^{+0.15}_{-0.21}$ & $1.75^{+0.25}_{-0.24}$ & $42.92^{+0.02}_{-0.08}$ & $-13.37^{+0.03}_{-0.07}$ & $181/176$ \\
        
        NuSTARJ204020-0056.1$^{\mathrm{b}}$&Borus & $23.71^{+0.18}_{-0.20}$ & $2.13^{+0.22}_{-0.60}$ & $44.00^{+0.04}_{-0.10}$ & $-12.98^{+0.03}_{-0.09}$ & $132/134$ \\
        NuSTARJ211517+0608.7$^{\mathrm{a}}$&Borus & $23.69^{+0.09}_{-0.10}$ & $1.99^{+0.23}_{-0.23}$ & $43.33^{+0.03}_{-0.04}$ & $-13.30^{+0.01}_{-0.14}$ & $166/190$ \\
        NuSTARJ213900-2640.4 &Borus & $22.54^{+0.25}_{-0.25}$ & $1.76^{+0.13}_{-0.11}$ & $44.40^{+0.01}_{-0.04}$ & $-12.65^{+0.01}_{-0.04}$ & $154/166$ \\
        NuSTARJ221133+1841.5 &Borus & $22.55^{+0.17}_{-0.23}$ & $1.97^{+0.19}_{-0.19}$ & $44.12^{+0.01}_{-0.17}$ & $-12.97^{+0.01}_{-0.25}$ & $166/175$ \\

\end{longtable}

\twocolumn

\subsection{NSS80 sample distribution}
\label{sec: sample discussion}

\begin{figure}
	\includegraphics[width=\columnwidth]{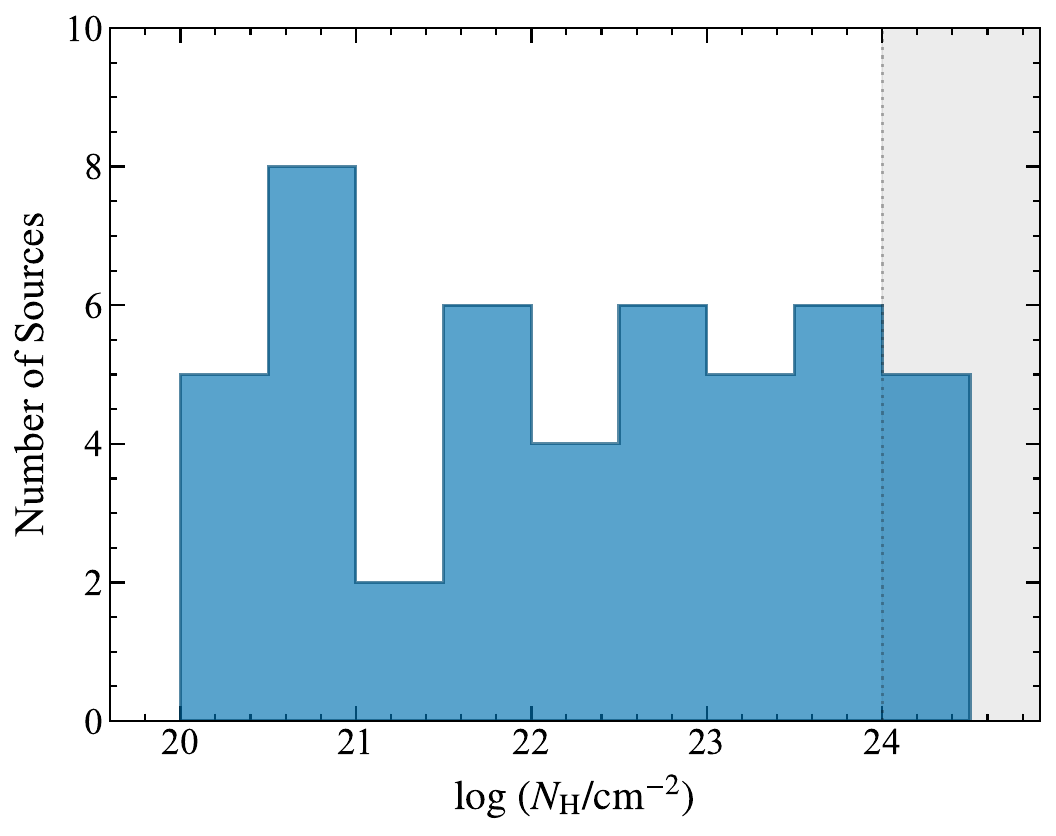}
    \caption{The column density ($\log(N_\mathrm{H}$/cm$^{-2})$) distribution of the NSS80 sources determined from spectral fitting using either the Power-law or Borus models. Bins have a width of 0.5. The grey region denotes CT AGN, in which we find 5 sources within the NSS80 sample.}
    \label{fig:Bright Sample Column Distribution}
\end{figure}

The NSS80 sample distribution of $\log(N_\mathrm{H}$/cm$^{-2})$ from the spectral fitting can be seen in Fig.\ref{fig:Bright Sample Column Distribution}. The NSS80 sample has a $\log(N_\mathrm{H}$/cm$^{-2})$ range of $20.2 \leq \log(N_\mathrm{H}$/cm$^{-2})\leq 24.38$, which lies within the range of the \citetalias{pfeifle2022bass} sample range of $20.18\leq\log(N_\mathrm{H}$/cm$^{-2})\leq25.40$ (excluding their upper bound sources) and the \citetalias{Asmus_diagnostic} sample of $19\leq\log(N_\mathrm{H}$/cm$^{-2})\leq24.70$ (excluding upper and lower bound sources). We find the NSS80 sample does lack extreme CT sources unlike the \textit{Swift/BAT} samples, potentially due to the high flux cut applied to the NSS80 candidate sources, in addition to the significantly smaller sample size by a factor of $\sim10$ compared to \citetalias{pfeifle2022bass}. The NSS80 sample includes a significantly more luminous range of AGN compared to both \citetalias{Asmus_diagnostic} and \citetalias{pfeifle2022bass}, with a median X-ray luminosity of $\langle\log \left( L_{2-10\mathrm{keV}} / \mathrm{erg\ s}^{-1} \right)\rangle= 43.50$. The similar distribution of sources in column density between the samples can provide a good comparison between the X-ray-MIR diagnostic fitting and the CT fraction in later sections.

\section{Fitting the X-ray-MIR Diagnostic }
\label{sec: The X-ray-MIR diagnostic}

Here we briefly discuss the K-correction applied to the \textit{WISE} MIR magnitudes for the NSS80 sample, before describing the Monte Carlo methods used to fit the NSS80 sample to the X-ray-MIR diagnostic in section~\ref{sec: Sample fitting methods}.

\subsection{Sample fitting methods}
\label{sec: Sample fitting methods}

Before fitting the NSS80 sample to the X-ray-MIR diagnostic, we first applied a K-correction to the \textit{WISE} magnitudes to account for the effect of redshift. A more detailed description of our methods and approximations are discussed in Appendix~\ref{sec:K-correction}.

Monte Carlo (MC) methods were used to fit the NSS80 sample to the X-ray-MIR diagnostic equation, following the methodology of \citetalias{pfeifle2022bass}. We did so to prevent any differences in the final NSS80 and \citetalias{pfeifle2022bass} X-ray-MIR diagnostic fits to be caused by the methods used.

The NSS80 sample was bootstrapped by taking a random selection of 47 sources from the the full sample, including repeats. This method accounts for the large intrinsic scatter of the sample itself. 

In each repeat, to account for the uncertainties in both the column density ($\log(N_\mathrm{H}$/cm$^{-2})$) and the luminosity ratio ($\log\left( {L}_{2-10\mathrm{keV}} / {L}_{12\mathrm{\mu m}}\right)$ a new point for each source was generated from a Gaussian distribution of the respective uncertainties. Due to asymmetrical uncertainties in both the luminosity ratio and the column density, the Gaussian distribution was taken separately for the upper and lower uncertainties. The upper-bound sources here are approximated to have a uniform distribution to $\log(N_\mathrm{H}$/cm$^{-2}) = 19$, in accordance to the methods of \citetalias{pfeifle2022bass}.

Orthogonal Distance Regression (ODR) was then applied to the sampled data using \texttt{SCIPY ODR} \citep{VirtanenSCIPY}. ODR is used to fit the X-ray-MIR diagnostic model in the form of equation~\ref{eqn: default model}, while accounting for both the luminosity ratio and column density uncertainties (with the approximation of symmetry). The fitting process was then repeated with a new bootstrap sample for a total of 10,000 times. The best fit was selected as the 50th percentile values with uncertainty from the 16th to 84th percentile range \citepalias{pfeifle2022bass}.

\begin{equation}
\label{eqn: default model}
    \log\left(\frac{L_{2-10\mathrm{keV}}}{  L_{12\mathrm{\mu m}}}\right) = a + b\times 10^{c(\log(N_\mathrm{H}) - 20)}
\end{equation}

Sources with upper bounds in $\log(N_\mathrm{H}$/cm$^{-2}))$ were kept in the sample for MC fitting, to improve the convergence of the parameters. Removing the upper bound sources resulting in inconsistent parameters from fitting. We also applied a different prior to the methods of \citetalias{pfeifle2022bass}. We used a uniform distribution of the upper bound sources given by the model fit down to a column density of $\log(N_\mathrm{H}$/cm$^{-2})=20.2$, corresponding to the lowest column density within the NSS80 sample. Changing the lower bound was done to have a value more representative of the sample, however, the change did not affect the fit significantly, so the decision was made to follow the methods of \citetalias{pfeifle2022bass} with a uniform distribution to $\log(N_\mathrm{H}$/cm$^{-2}) = 19$.

The inverse of equation~\ref{eqn: default model} was initially fit to the sample, but was found to not converge with any initial parameter values used in ODR as found in \citetalias{pfeifle2022bass} (see Appendix~\ref{app: inverse}). As such, equation~\ref{eqn: default model} was used for the MC fitting, and then inverted to predict the source column density. The ODR fitting realisations of the MC fitting for ($\log\left( {L}_{2-10\mathrm{keV}} / {L}_{12\mathrm{\mu m}}\right)$) can be seen in Fig.~\ref{fig:Project X-ray scatter}. A secondary X-ray-MIR diagnostic model from \citetalias{pfeifle2022bass} using $\log\left(L_{22\mathrm{\mu m}} /L_{4.6\mathrm{\mu m}}\right)$ was fit to the NSS80 sample, and found to be very poorly constrained. We therefore decided not to include the results within our main body of work, but discuss the fit to this purely MIR diagnostic in the Appendix (see section~\ref{app: inverse}).

The \citetalias{Asmus_diagnostic} sample was also fit to the X-ray-MIR diagnostic using the same methods of \citetalias{pfeifle2022bass} as a comparison to the NSS80 sample. Unlike the NSS80 or \citetalias{pfeifle2022bass} samples, the \citetalias{Asmus_diagnostic} sample contains lower bound sources in $\log(N_\mathrm{H}$/cm$^{-2})$ and upper bound sources in $\log({L}_{12\mathrm{\mu m}})$. The unbounded \citetalias{Asmus_diagnostic} sources were not consistent with the \citetalias{pfeifle2022bass} fitting methods\footnote{The \citetalias{pfeifle2022bass} sample does not include uncertainty measurements for $\log\left( {L}_{2-10\mathrm{keV}}/ \mathrm{erg\ s}^{-1}\right)$, and instead assumes uncertainties of $\Delta \log\left({L}_{2-10\mathrm{keV}}/ \mathrm{erg\ s}^{-1}\right)=\pm 0.1$ for $\log(N_\mathrm{H}$/cm$^{-2})<24$ and $\Delta \log\left({L}_{2-10\mathrm{keV}}/ \mathrm{erg\ s}^{-1}\right)=\pm 0.3$ for $\log(N_\mathrm{H}$/cm$^{-2})\geq24$.}, so to ensure fitting the three samples were consistent we removed the unbounded sources from the \citetalias{Asmus_diagnostic} sample. It should be noted that 81 of the \citetalias{pfeifle2022bass} sources are also contained within the full \citetalias{Asmus_diagnostic} sample, with largely consistent measurements within uncertainty in luminosity and obscuration.

The NSS80 and \citetalias{pfeifle2022bass} samples were also fit to a linear diagnostic model similar as in \citetalias{Asmus_diagnostic}, modified to use the luminosity ratio rather than the flux ratio. To do so, we followed the same MC and bootstrapping methods as described above. The samples were fit to equation~\ref{eqn: linear model}, with the \lq $b$\rq$ $ parameter fixed to be positive following the shape of \citetalias{Asmus_diagnostic}. 

\begin{equation}
\label{eqn: linear model}
    \log\left(\frac{N_{\mathrm{H}}}{10^{22.8} \mathrm{cm^{-2}} }\right) = a + b\times \log\left(\frac{L_{2-10\mathrm{keV}}}{  L_{12\mathrm{\mu m}}}\right)
\end{equation}

Equation~\ref{eqn: linear model} limits the fit to column densities to $\log(N_\mathrm{H}$/cm$^{-2}))>22.8$ to fit the diagnostic model to only obscured sources \citepalias{Asmus_diagnostic}. For the NSS80 this left 18 sources for the fit, with 171 for the \citetalias{pfeifle2022bass} sample.

\section{Analysis of the X-ray-MIR diagnostic}
\label{sec: mir results}

\begin{figure*}
	\includegraphics[width=\textwidth]{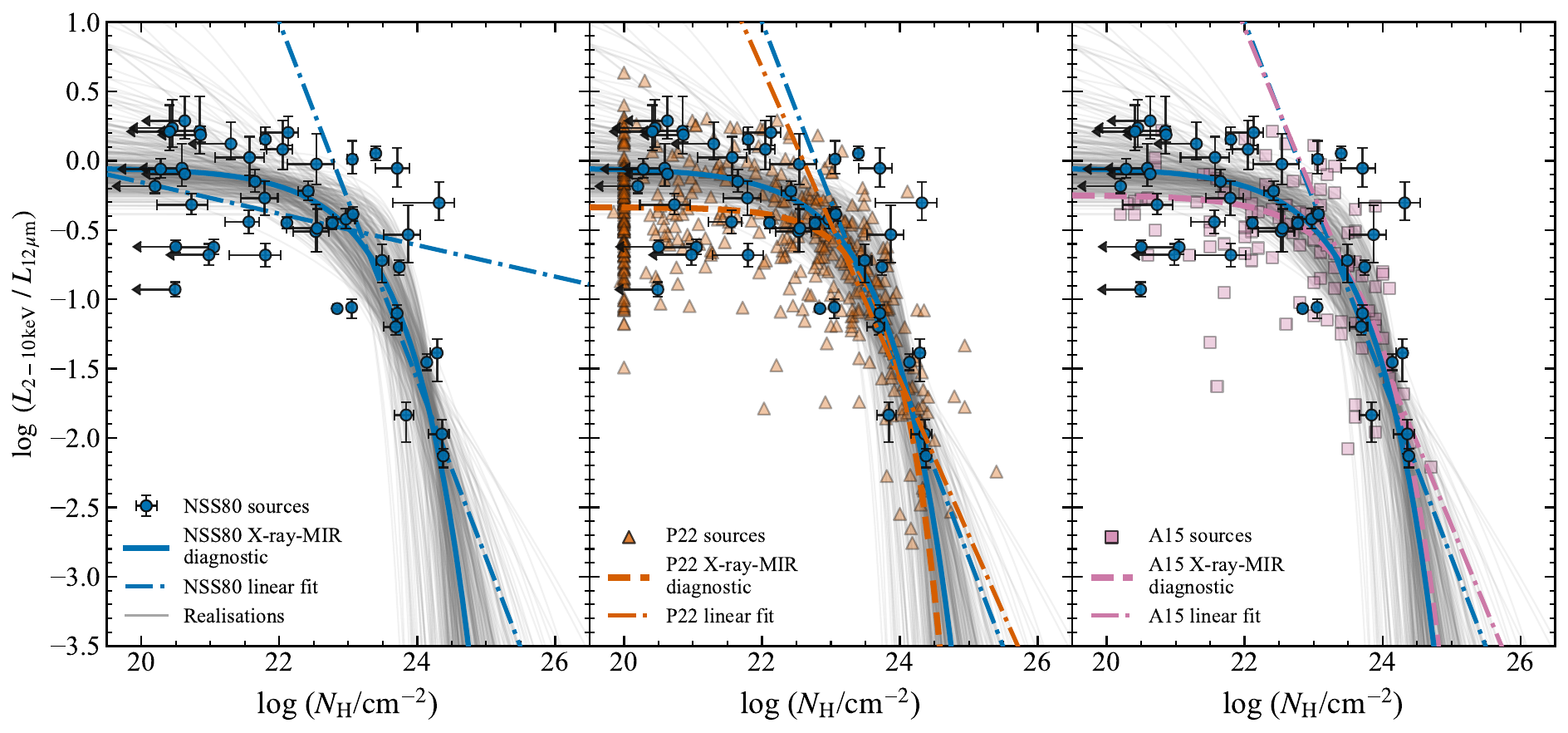}
    \caption{The logarithmic luminosity ratio $\log\left( {L}_{2-10\mathrm{keV}} / {L}_{12\mathrm{\mu m}}\right)$ against column density $\log(N_\mathrm{H}$/cm$^{-2})$ for the NSS80 (blue circles), \citetalias{pfeifle2022bass} (orange triangles; centre) and \citetalias{Asmus_diagnostic} (pink squares; right) samples. We plot the NSS80 X-ray-MIR diagnostic (blue line) and a sample of 300 of the MC fit realisations (grey lines) compared to the best fit of the \citetalias{pfeifle2022bass} (orange dashed curve) and \citetalias{Asmus_diagnostic} (pink dashed curve). Dot-dashed lines show the linear fit close to the form determined by \citetalias{Asmus_diagnostic} for obscured sources $\log(N_\mathrm{H}$/cm$^{-2})\geq22.8$ and (for NSS80) unobscured AGN with $\log(N_\mathrm{H}$/cm$^{-2})<22.8$. The NSS80 X-ray luminosity and respective uncertainty is determined from the XSPEC Lumin function through spectral fitting. All other luminosities are taken from the respective catalogues of \citetalias{pfeifle2022bass} and \citetalias{Asmus_diagnostic}. Sources in the NSS80 sample with an upper bound in column density are plotted using an arrow with the data point set at 1$\sigma$ uncertainty. An initial offset in the NSS80 X-ray-MIR diagnostic curve relative to the \citetalias{pfeifle2022bass} and \citetalias{Asmus_diagnostic} sample curves can be seen in the centre and right panels.}
    \label{fig:Project X-ray scatter}
\end{figure*}

The resulting NSS80 sample fit of the X-ray-MIR diagnostic ($\log\left( {L}_{2-10\mathrm{keV}} / {L}_{12\mathrm{\mu m}}\right)$) is shown in Table~\ref{tab:Parameters} and Fig.~\ref{fig:Project X-ray scatter}.

\begin{table}
\renewcommand{\arraystretch}{1.5}
	\centering
	\caption{Parameters for the NSS80, \citetalias{pfeifle2022bass} and \citetalias{Asmus_diagnostic} model fits. The parameters correspond to equation~\ref{eqn: default model} or equation~\ref{eqn: linear model} (for the linear models).}
	\label{tab:Parameters}
	\begin{tabular}{lccc} 
		\hline
		Model & $a$ & $b$ & $c$\\
		\hline
        NSS80 $L_{2-10\mathrm{keV}} / L_{12\mathrm{\mu m}}$ & $-0.05^{+0.39}_{-0.13}$ & $-0.012^{+0.012}_{-0.14}$ & $0.52^{+0.46}_{-0.25}$\\
        
        NSS80 Linear
        & $-0.02^{+0.31}_{-0.48}$ & $-0.78^{+0.32}_{-0.20}$ & -\\
        
		\citetalias{pfeifle2022bass} $L_{2-10\mathrm{keV}} / L_{12\mathrm{\mu m}}$ & $-0.34^{+0.06}_{-0.06}$ & $-0.003^{+0.002}_{-0.005}$ & $0.62^{+0.13}_{-0.12}$\\

        \citetalias{pfeifle2022bass} Linear & $-0.21^{+0.08}_{-0.09}$ & $-0.89^{+0.09}_{-0.07}$ & -\\
          
        \citetalias{Asmus_diagnostic} $L_{2-10\mathrm{keV}} / L_{12\mathrm{\mu m}}$ & $-0.25^{+0.32}_{-0.10}$ & $-0.006^{+0.005}_{-0.11}$ & $0.58^{+0.31}_{-0.31}$\\
        
        \citetalias{Asmus_diagnostic} Linear & $-0.00^{+0.13}_{-0.13}$ & $-0.84^{+0.16}_{-0.13}$ & -\\

        Combined $L_{2-10\mathrm{keV}} / L_{12\mathrm{\mu m}}$ & $-0.11^{+0.11}_{-0.08}$ & $-0.012^{+0.010}_{-0.031}$ & $0.52^{+0.22}_{-0.13}$\\

		\hline
	\end{tabular}
\end{table}

\begin{figure*}
    \includegraphics[width=0.9\textwidth]{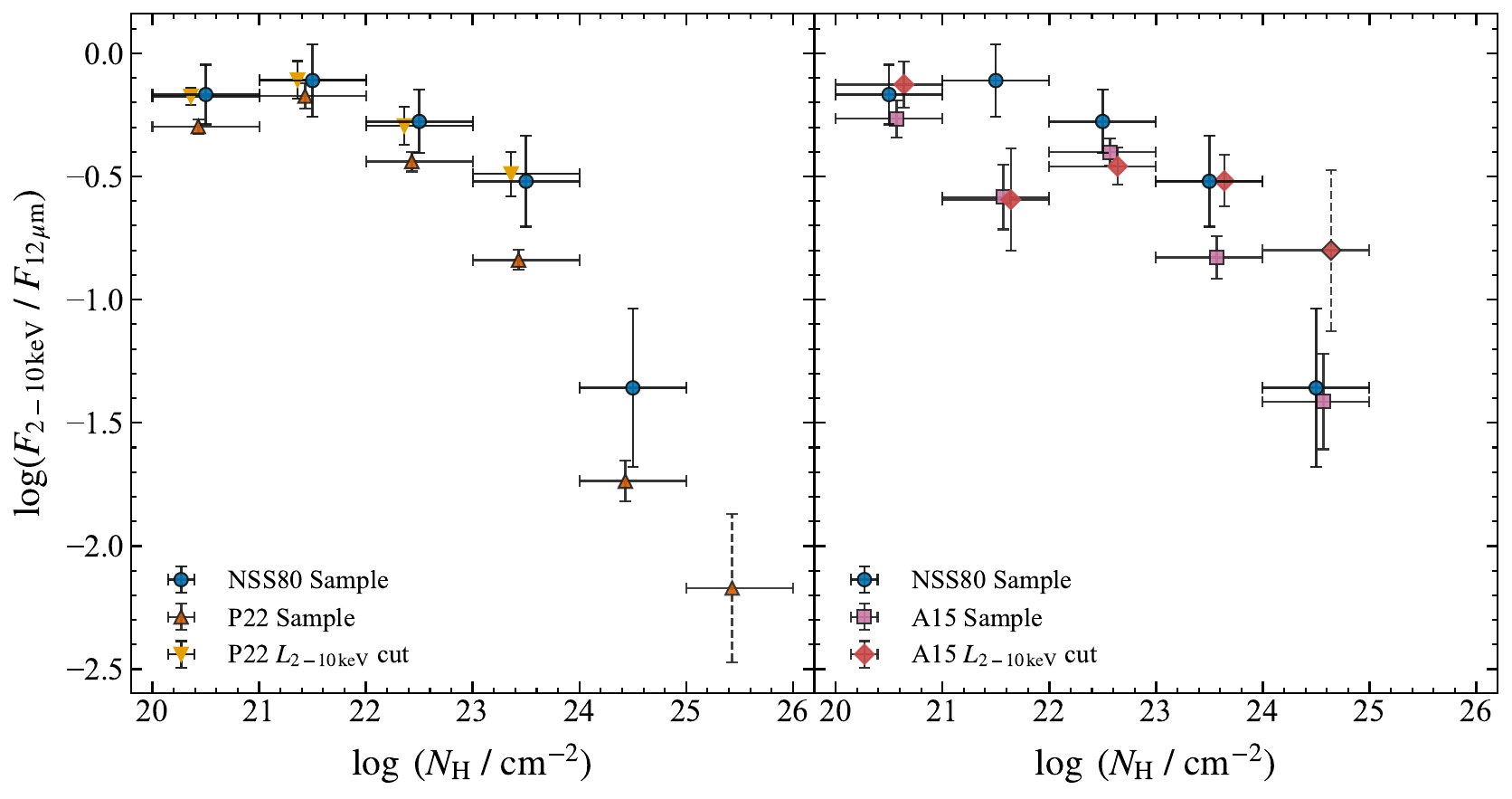}
    \caption{The distribution of $\log\left( F_{2-10\mathrm{keV}} / F_{12\mathrm{\mu m}}\right)$ plotted as a function of column density ($\log(N_\mathrm{H}$/cm$^{-2})$) with a bin width of 1. The NSS80 sample (blue circles) is compared to (left) the \citetalias{pfeifle2022bass} sample of \textit{Swift/BAT} AGN (orange triangles) and (right) the \citetalias{Asmus_diagnostic} sample of local AGN (pink squares). A comparison is also made with a luminosity cut of $\log \left( L_{2-10\mathrm{keV}} / \mathrm{erg\ s}^{-1} \right)\geq 43.30$ applied to \citetalias{pfeifle2022bass} (yellow inverted triangle) and $\log \left( L_{2-10\mathrm{keV}} / \mathrm{erg\ s}^{-1} \right)\geq 42.70$ applied to \citetalias{Asmus_diagnostic} (red diamond) to better match the luminosities of the NSS80 sources. The uncertainty is provided as the standard error of each respective bin except the bins with vertical dashed error bars, which include only one source. For these bins we instead use the uncertainty on the flux ratio for the given source. All bins for \citetalias{pfeifle2022bass} and \citetalias{Asmus_diagnostic} have been offset for clarity. Note for the NSS80 sample the X-ray flux was calculated using the XSPEC Flux function, with the \citetalias{Asmus_diagnostic} sample calculated from the provided luminosity, and \citetalias{pfeifle2022bass} from the provided flux measurements within the respective catalogues. We observe the luminosity cut subsample of \citetalias{pfeifle2022bass} follows more closely the trend of the NSS80 sample, implying the separation of samples is X-ray luminosity dependant.}
    \label{fig:X-ray binned flux}

\end{figure*}

We theorised that the large uncertainties of the NSS80 sample fit (here-on the NSS80 X-ray-MIR diagnostic) can be partly attributed to the low number of sources, which is evidenced by the larger standard error of the NSS80 sample relative to the \citetalias{pfeifle2022bass} sample seen in Fig.~\ref{fig:X-ray binned flux}. We note however, that the uncertainty of the NSS80 X-ray-MIR diagnostic is similar to the fit of the \citetalias{Asmus_diagnostic} sample, which has over twice as many sources. It is therefore likely that other factors including the sample selection methods also have an influence in the uncertainty of the X-ray-MIR diagnostic fits. The scatter of the three samples are all similar, with a close distribution of sources for the NSS80 and the \citetalias{pfeifle2022bass} or \citetalias{Asmus_diagnostic} sample in Fig.~\ref{fig:Project X-ray scatter}.

Fig.~\ref{fig:Project X-ray scatter} shows the distribution of the NSS80 sample, with a comparison to the \citetalias{pfeifle2022bass} (centre panel) and \citetalias{Asmus_diagnostic} sample (right panel). The large uncertainty in the MC fit for the NSS80 sample is shown by the distribution of the realisations, plotted from the bootstrap ODR fitting. The X-ray-MIR diagnostic fit of the \citetalias{Asmus_diagnostic} and \citetalias{pfeifle2022bass} sample can be seen initially lower than the NSS80 X-ray-MIR diagnostic, with all samples showing the same trend at higher obscuration.

As seen in Table~\ref{tab:Parameters}, the parameters \lq $b$\rq$ $ and \lq$c$\rq$ $ (equation \ref{eqn: default model}) are consistent within uncertainty between \citetalias{Asmus_diagnostic} and \citetalias{pfeifle2022bass} samples in $\log\left( {L}_{2-10\mathrm{keV}} / {L}_{12\mathrm{\mu m}}\right)$. The NSS80 sample however has a greater \lq $a$\rq$ $ parameter, than the \textit{Swift/BAT} samples, suggesting an offset in $\log\left( {L}_{2-10\mathrm{keV}} / {L}_{12\mathrm{\mu m}}\right)$ for a given column density. This offset is evidenced in Fig.~\ref{fig:Project X-ray scatter}, in which the NSS80 sources are on average above the \citetalias{Asmus_diagnostic} and \citetalias{pfeifle2022bass} prediction curve (dashed lines).

Despite this offset, the overall shape of the curve is seen to be the same, implied by the agreement in the power parameter (\lq$c$\rq) and from the similar trends observed in Fig.~\ref{fig:Project X-ray scatter}, with a close shape at high obscuration ($\log(N_\mathrm{H}$/cm$^{-2})\gtrsim23.5$). The NSS80 sample also lies qualitatively within the range of \citetalias{Asmus_diagnostic} and \citetalias{pfeifle2022bass} for both Fig.~\ref{fig:Project X-ray scatter}, suggesting that the AGN samples are similar, and are therefore a good comparison between redshifts.

The offset in the NSS80 MC fit can be attributed to a higher observed X-ray luminosity in the sample, compared to both \citetalias{pfeifle2022bass} and \citetalias{Asmus_diagnostic}. To investigate the offset in $\log\left( {L}_{2-10\mathrm{keV}} / {L}_{12\mathrm{\mu m}}\right)$ between the NSS80 and the \textit{Swift/BAT} samples, the flux ratio was used to compare the measurements from both samples as seen in Fig.~\ref{fig:X-ray binned flux}. The sources are binned by $\log(N_\mathrm{H}$/cm$^{-2})$ with a bin width of 1. Fig.~\ref{fig:X-ray binned flux} shows the same decrease in ratio as Fig.~\ref{fig:Project X-ray scatter}, with the bins between the NSS80 and \citetalias{pfeifle2022bass} in good agreement for unobscured sources, but then diverge as $\log(N_\mathrm{H}$/cm$^{-2})$ increases, suggesting a difference in $\log\left( {L}_{2-10\mathrm{keV}} / {L}_{12\mathrm{\mu m}}\right)$ distribution between the samples.

A luminosity cut was applied to the \citetalias{pfeifle2022bass} and \citetalias{Asmus_diagnostic} samples to increase their median X-ray luminosities to be comparable with the NSS80 sample, which has $\langle\log \left( L_{2-10\mathrm{keV}} / \mathrm{erg\ s}^{-1} \right)\rangle= 43.50$. The results of this can be seen in Fig.~\ref{fig:X-ray binned flux}. The minimum sample luminosity was applied at $\log \left( L_{2-10\mathrm{keV}} / \mathrm{erg\ s}^{-1} \right)\geq 42.70$ for the \citetalias{Asmus_diagnostic} sample and $\log \left( L_{2-10\mathrm{keV}} / \mathrm{erg\ s}^{-1} \right)\geq 43.30$ for the \citetalias{pfeifle2022bass} sample. The X-ray luminosity cuts increased the median values for the two samples to $\langle \log \left( L_{2-10\mathrm{keV}} / \mathrm{erg\ s}^{-1} \right)\rangle= 43.52$ and $\langle\log \left( L_{2-10\mathrm{keV}} / \mathrm{erg\ s}^{-1} \right)\rangle= 43.62$ for \citetalias{Asmus_diagnostic} and \citetalias{pfeifle2022bass} respectively. We found the heavily obscured column density distribution ($23\leq\log(N_\mathrm{H}$/cm$^{-2})\leq25$) of the luminosity cut \citetalias{pfeifle2022bass} sample to be statistically different to the NSS80 sample ($p=.002$) using an Anderson-Darling (AD) test. We interpreted the X-ray luminosity cut as an effective $N_{\mathrm{H}}$ cut which removed most of the heavily obscured AGN from the \citetalias{pfeifle2022bass} sample, which can also be seen in Fig.~\ref{fig:X-ray binned flux}. A more detailed analysis of the statistical tests of the X-ray luminosity and flux cut is provided in Appendix~\ref{Appendix: Analysis of the Luminosity Ratio}.

\subsection{The high X-ray luminosity X-ray-MIR diagnostic}
\label{sec: high Lx X-ray-MIR diagnostic}

We assessed the claim that the high average X-ray luminosity explains the initial offset seen in the NSS80 X-ray-MIR diagnostic relative to the \citetalias{Asmus_diagnostic} and \citetalias{pfeifle2022bass} diagnostic. To do so we combined the samples with the full NSS80 sample, and the luminosity cut \citetalias{pfeifle2022bass} and \citetalias{Asmus_diagnostic} samples (see Fig.~\ref{fig:X-ray binned flux}). Merging the samples was done in order to test the consistency of the $L_X$ cuts across samples, and provide a larger number of sources to reduce the uncertainty of the combined X-ray-MIR diagnostic fit. We initially removed the \citetalias{pfeifle2022bass} sources with $\log(N_\mathrm{H}$/cm$^{-2})=20$, since the value was not found by statistical fitting and instead chosen as a reasonable estimate of lower bound. We identified 81 sources that matched between \citetalias{Asmus_diagnostic} and \citetalias{pfeifle2022bass}. To prevent an overlap when fitting, we prioritised the \citetalias{Asmus_diagnostic} measurement when both were available, due to the more realistic errors on $\log(N_\mathrm{H}$/cm$^{-2})$, and the provided $L_{\mathrm{X}}$ errors. The \citetalias{pfeifle2022bass} measurement would be used if \citetalias{Asmus_diagnostic} provided a lower (or upper) bound for that source, to ensure the MC fitting methods remained consistent. Following the selection criteria, we were left with a combined sample of 179 sources.

\begin{figure}
	\includegraphics[width=\columnwidth]{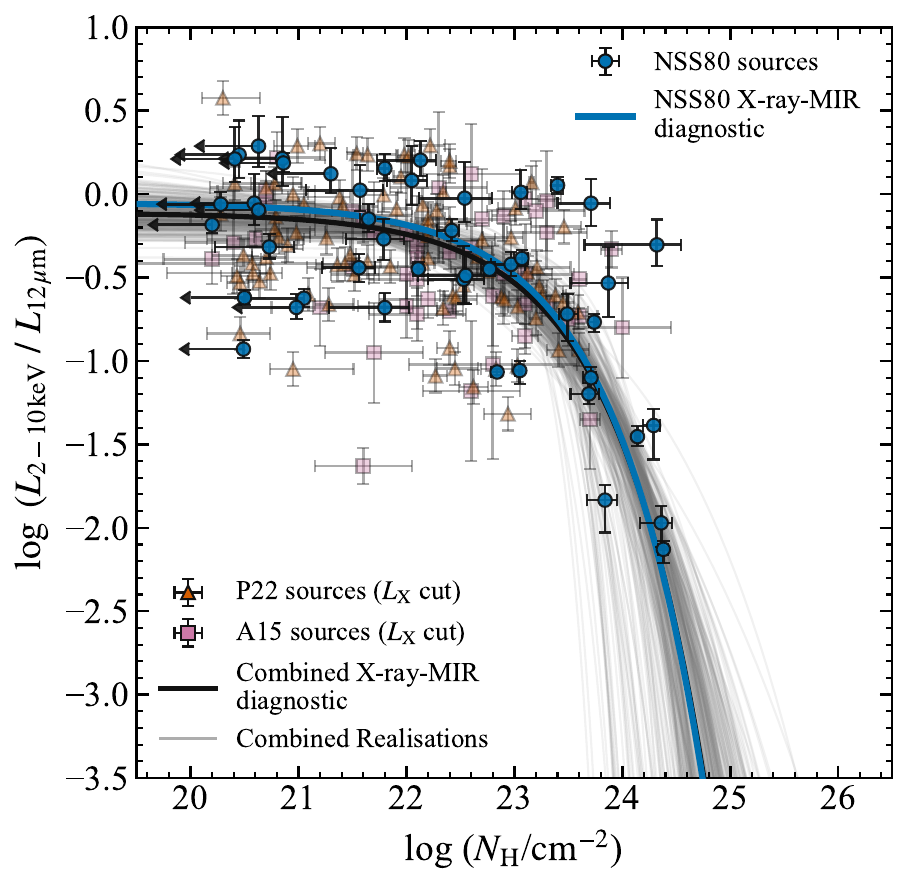}
    \caption{The luminosity ratio $\log\left( {L}_{2-10\mathrm{keV}} / {L}_{12\mathrm{\mu m}}\right)$ against column density $\log(N_\mathrm{H}$/cm$^{-2})$ of the NSS80 sample with a high luminosity cut of the \citetalias{pfeifle2022bass} (orange triangles) and \citetalias{Asmus_diagnostic} (pink squares) samples at $\log \left( L_{2-10\mathrm{keV}} / \mathrm{erg\ s}^{-1} \right)\geq 43.30$ and $\log \left( L_{2-10\mathrm{keV}} / \mathrm{erg\ s}^{-1} \right)\geq 42.70$ respectively (see Fig.~\ref{fig:X-ray binned flux}). We plot the X-ray-MIR diagnostic of the combined samples (black line) with a sample of 300 MC fit realisations (grey lines) compared to the NSS80 X-ray-MIR diagnostic (blue line). The NSS80 X-ray luminosity was calculated using the XSPEC Lumin function, with \citetalias{Asmus_diagnostic} and \citetalias{pfeifle2022bass} luminosity measurements from the respective catalogues. Sources with an upper bound in $\log(N_\mathrm{H}$/cm$^{-2})$ are plotted using an arrow with the data point set at 1$\sigma$ uncertainty. We note there is close agreement between the curve shapes, suggesting the initial offset in curves (see Fig.~\ref{fig:Project X-ray scatter}) was dependant on the X-ray luminosity.}
    \label{fig:Combined X-ray-MIR diagnostic plot}
\end{figure}

We fit the combined sample with the same methods as described in section~\ref{sec: Sample fitting methods}, with the combined X-ray-MIR diagnostic shown in Fig.~\ref{fig:Combined X-ray-MIR diagnostic plot}. It can be seen from Fig.~\ref{fig:Combined X-ray-MIR diagnostic plot} and from Table~\ref{tab:Parameters} that the combined X-ray-MIR diagnostic is in close agreement with the NSS80 X-ray-MIR diagnostic. The similarity of both curves further implies that a higher average X-ray luminosity causes the higher initial offset from the NSS80. The larger sample size can also be seen with the realisations to significantly reduce the uncertainty in the X-ray-MIR diagnostic relative to the NSS80 sample up to the obscured region ($\log(N_\mathrm{H}$/cm$^{-2})<23$; see Table~\ref{tab:Parameters} and Fig.~\ref{fig:Combined X-ray-MIR diagnostic plot}).

Utilising the X-ray-MIR diagnostic on a new catalogue or sample in the future will therefore require an understanding of the distribution of the X-ray luminosity. More luminous sources might have a larger \lq$a$\rq\ term in the model equation (equation \ref{eqn: default model}). The NSS80 X-ray-MIR diagnostic obeys a similar curve shape to the provided \citetalias{pfeifle2022bass} X-ray-MIR diagnostic and the \citetalias{Asmus_diagnostic} X-ray-MIR diagnostic with \textit{Swift/BAT} (see Fig.~\ref{fig:Project X-ray scatter}). It is implied therefore that the X-ray-MIR diagnostic can still be applied to higher redshift AGN but that the properties of the sample must be taken into account.

A higher average accretion rate between the NSS80 and the \textit{Swift/BAT} samples would result in the observed luminosity offset with an increased relative X-ray luminosity. As the accretion disk is the primary power source of the AGN \citep{ALEXANDER2025DecadeReview}, a greater accretion rate onto the SMBH would increase the disk luminosity, and therefore the X-ray luminosity through inverse Compton scattering \citep{Eddington_xray}. The higher accretion rate would in turn decrease the relative MIR luminosity as the dust grains within the torus and polar outflows are sublimated and expelled by radiation pressure (\citetalias{Asmus_diagnostic}; \citealp{Hoenig_2019}). The expulsion of dust grains further decreases the covering factor and obscuration from the torus, and therefore the MIR emission \citetalias{Asmus_diagnostic}. As a result the $\log\left( L_{2-10\mathrm{keV}} / L_{12\mathrm{\mu m}}\right)$ increases for a given $\log(N_\mathrm{H}$/cm$^{-2})$.  We note however that the \lq receding torus model\rq$ $ described \citep{Lawrence_1991,Simpson_2005} is in contradiction with the work of \citet{Horst_2008} and \citet{Wada_2015}, with the latter suggesting that the covering fraction increases with increasing intrinsic X-ray luminosity. The increased covering factor in turn increases the MIR emission from the torus and decreases the observed X-ray emission, which decreases the luminosity ratio. To determine which torus model best describes the observed trend in $\log\left( {L}_{2-10\mathrm{keV}} / {L}_{12\mathrm{\mu m}}\right)$, the Eddington luminosity (L$_{Edd}$) would be needed to compare the accretion rates for the NSS80 and \textit{Swift/BAT} samples \citep{Eddington_xray}. This provides grounds for future research.

\subsection{The Linear diagnostic method}
\label{sec: linear diagnostic}

We fit the NSS80 and \citetalias{pfeifle2022bass} sample to the estimation method used in \citetalias{Asmus_diagnostic} (see equation~\ref{eqn: linear model}), with the resulting fits seen in Fig.~\ref{fig:Project X-ray scatter}. Due to the different fitting methods, the conversion of the equation to the luminosity ratio (instead of flux) and the source removal of the \citetalias{Asmus_diagnostic} sample (see section~\ref{sec: Sample fitting methods}) we re-fit the linear diagnostic equation to the \citetalias{Asmus_diagnostic} sample. We found the new fit within this paper obeys a close trend to the model and parameters in \citetalias{Asmus_diagnostic}, but with a shallower slope. We attribute the difference to the provided reasons above, and will continue to use the new model as a comparison to the NSS80 and \citetalias{pfeifle2022bass} samples to remove the impact of fitting methods in our analysis.

We find that all three modelled estimation fits are in agreement with each other considering uncertainty. As the \citetalias{Asmus_diagnostic} equation only considers the obscured sources ($\log(N_\mathrm{H}$/cm$^{-2})\geq22.8$), the best fit is not as heavily skewed by the higher luminosity offset observed in the NSS80 sample. The \citetalias{Asmus_diagnostic} equation further emphasises the same trend in luminosity ratio is observed at higher redshifts.

We included a further linear fit for the NSS80 sample for unobscured sources. Fig.~\ref{fig:Project X-ray scatter} (left) shows the NSS80 for unobscured sources up to $\log(N_\mathrm{H}$/cm$^{-2})=22.8$ (blue dash-dot). We observe a slight negative relation with increasing column density implying some X-ray absorption with increasing column density, even at low obscuration. The X-ray-MIR diagnostic fits of all three samples at $\log(N_\mathrm{H}$/cm$^{-2})\geq22.8$ show a slope, though the unobscured linear fit does not agree with the corresponding luminosity ratio. The disagreement could be due to the large uncertainty on fitting at low obscuration \citep{Masini_2018}, with large scatter among the few sources in the NSS80 sources. We found that increasing the limit for unobscured sources up to $\log(N_\mathrm{H}$/cm$^{-2})=23.1$ resulted in a steeper negative gradient, in agreement with the X-ray-MIR diagnostic showing a decreasing luminosity ratio for increasing obscuration.

\subsubsection{The extent of host galaxy contamination}
\label{sec: The host galaxy contamination}

In this work we assume that the impact of using total galaxy MIR emission is minimal. \citetalias{Asmus_diagnostic} separates nuclear emission in their work, and by comparing with this we assess the contribution of non-nuclear emission to the total to be low. While we cannot eliminate the potential for non-negligible contribution from the host galaxy, the agreement of both parameters in the linear diagnostic equation fit of the NSS80 and \citetalias{Asmus_diagnostic} samples imply that the \citetalias{Asmus_diagnostic} relation for obscured AGN remains consistent, even when including host galaxy MIR emission for the NSS80 and \citetalias{pfeifle2022bass} sample. A correction to the \textit{WISE} MIR emission was considered for the NSS80 sources, but was decided against as such corrections are highly uncertain and not recommended for individual objects \citep{cor_asmus_2020}. The correction used in \citet{cor_asmus_2020} was measured in the local universe, so an application to the NSS80 sources may also be invalid. The impact of contamination was discussed in \citetalias{pfeifle2022bass}, where they found the X-ray-MIR diagnostic can still be used with contaminated sources.

Less dominant contamination from the host galaxy in the NSS80 sample relative to the \citetalias{pfeifle2022bass} sample does not fully account for the offset in luminosity ratio. Heavy contamination from star-formation in the host galaxy increases the 12 $\mathrm{\mu m}$ luminosity, and therefore decreases $\log\left( L_{2-10\mathrm{keV}} / L_{12\mathrm{\mu m}}\right)$. \citetalias{pfeifle2022bass} determined using the AGN emission without contamination resulted in an increase of $\Delta\log\left( {L}_{2-10\mathrm{keV}} / {L}_{12\mathrm{\mu m}}\right)\approx0.5$ for host galaxy dominant sources. The \citetalias{Asmus_diagnostic} sample shown in Fig.~\ref{fig:Project X-ray scatter} (right) uses the intrinsic 12 $\mathrm{\mu m}$ luminosity when fitting the X-ray-MIR diagnostic. We observe that the offset estimated by \citetalias{pfeifle2022bass} is significantly less than the range of the NSS80 sample, and find the initial offset of the NSS80 X-ray-MIR diagnostic to be well within the \citetalias{pfeifle2022bass} estimate relative to both the \citetalias{pfeifle2022bass} and \citetalias{Asmus_diagnostic} X-ray-MIR diagnostic. We therefore find that the difference is a result of a higher relative X-ray luminosity and not host galaxy contamination.

\subsection{The Compton thick probability}
\label{sec: The CT probability}

\begin{figure}
	\includegraphics[width=\columnwidth]{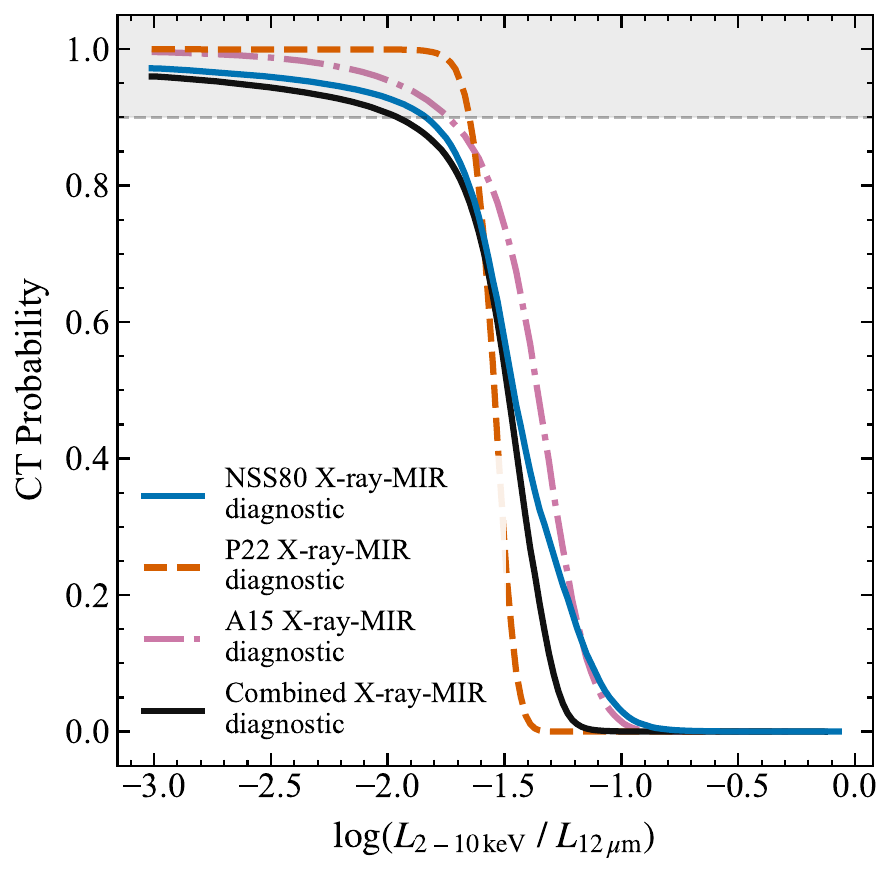}
    \caption{The probability for a source to be CT ($\log(N_\mathrm{H}$/cm$^{-2}))\geq24$) for a given luminosity ratio ($\log\left( {L}_{2-10\mathrm{keV}} / {L}_{12\mathrm{\mu m}}\right)$) using the distribution of bootstrap samples within bins of width 0.02. The blue curve denotes the CT probability from the model fit of the NSS80 X-ray-MIR diagnostic, while the orange dashed and pink dot-dashed curve denotes the CT probability of the \citetalias{pfeifle2022bass} and \citetalias{Asmus_diagnostic} X-ray-MIR diagnostic respectively. The CT probability of the combined X-ray-MIR diagnostic (black line; see Fig.~\ref{fig:Combined X-ray-MIR diagnostic plot}) has also been plotted, which shows a steeper distribution than the NSS80 sample. Confidence of a source $>90\%$ CT probability is indicated by the grey region.}
    \label{fig:CT probability realisations}
\end{figure}

The ODR fit realisations were used to model the distribution of the X-ray-MIR diagnostic fit on the sample, which can be used to calculate the probability of a source being CT provided a measurement of $\log\left( {L}_{2-10\mathrm{keV}} / {L}_{12\mathrm{\mu m}}\right)$. Counting the fraction of realisations within a given range that predict a source would be CT provides the CT probability for that luminosity ratio. Fig.~\ref{fig:CT probability realisations} shows the probability distribution for the NSS80, \citetalias{pfeifle2022bass}, the \citetalias{Asmus_diagnostic}, and combined X-ray-MIR diagnostic. 

It can be seen in Fig.~\ref{fig:CT probability realisations} that the slope of the NSS80 probability is significantly shallower than that of \citetalias{pfeifle2022bass}, which is consistent with the higher uncertainty in the NSS80 MC fit (see Table~\ref{tab:Parameters}). The probability curve from \citetalias{Asmus_diagnostic} shows a similar distribution to the NSS80 with a steeper gradient despite the difference in X-ray luminosity and redshift.

The probability distribution of the combined X-ray-MIR diagnostic is also shown in Fig.~\ref{fig:CT probability realisations}. The curve shows a significantly steeper curve than the NSS80 X-ray-MIR diagnostic due to the reduced uncertainty in the model fit. The combined X-ray-MIR diagnostic curve is in excellent agreement with the trend of the NSS80 X-ray-MIR diagnostic at lower luminosity ratios, which can also be interpreted from Fig.~\ref{fig:Combined X-ray-MIR diagnostic plot}. The slightly lower confidence in the CT probability for $\log\left( {L}_{2-10\mathrm{keV}} / {L}_{12\mathrm{\mu m}}\right)\lesssim-1.6$ is potentially due to the few additional \citetalias{Asmus_diagnostic} sources within the CT region increasing the spread of values, and therefore uncertainty within the fit. 

The NSS80 X-ray-MIR diagnostic implies a 90$\%$ confidence in a source being CT for $\log\left( {L}_{2-10\mathrm{keV}} / {L}_{12\mathrm{\mu m}}\right)<-1.83$, with a greater confidence for both \citetalias{Asmus_diagnostic} and \citetalias{pfeifle2022bass} at 92$\%$ and 99$\%$ respectively. The confidence from the combined X-ray-MIR diagnostic is slightly lower with 90$\%$ confidence at $\log\left( {L}_{2-10\mathrm{keV}} / {L}_{12\mathrm{\mu m}}\right)<-1.95$. The method of using the CT probability allows for the quick identification of a likely CT source without a prediction of the column density, similar to the band ratio or hardness ratio. We emphasise however, that using Fig.~\ref{fig:CT probability realisations} requires an understanding of the luminosity distribution of the sample, to ensure an accurate probability.

\section{Application of the X-ray-MIR diagnostic to the wider NSS80 catalogue}

Here we apply the NSS80 X-ray-MIR diagnostic to the wider NSS80 catalogue. In section~\ref{sec: Catalogue reduction} we discuss the selection criteria necessary to apply the diagnostic to the catalogue. We then compare X-ray-MIR diagnostic predicted CT sources to the Band Ratio in section~\ref{sec: Application to the NSS80}. Finally we determine a statistical measure of the CT fraction within the NSS80 catalogue, and assess the result against previous research in section~\ref{sec: CT fraction analysis}.

\subsection{NSS80 catalogue source selection}
\label{sec: Catalogue reduction}

Following the work in validating the X-ray-MIR diagnostic in section~\ref{sec: mir results}, we can now apply this diagnostic to the wider NSS80 catalogue, where the majority of the sources are too X-ray faint for detailed X-ray spectral fitting. The application to the NSS80 aims to measure the observed CT fraction of the catalogue in a statistical way, and compare it to current literature values.

\begin{figure}
	\includegraphics[width=\columnwidth]{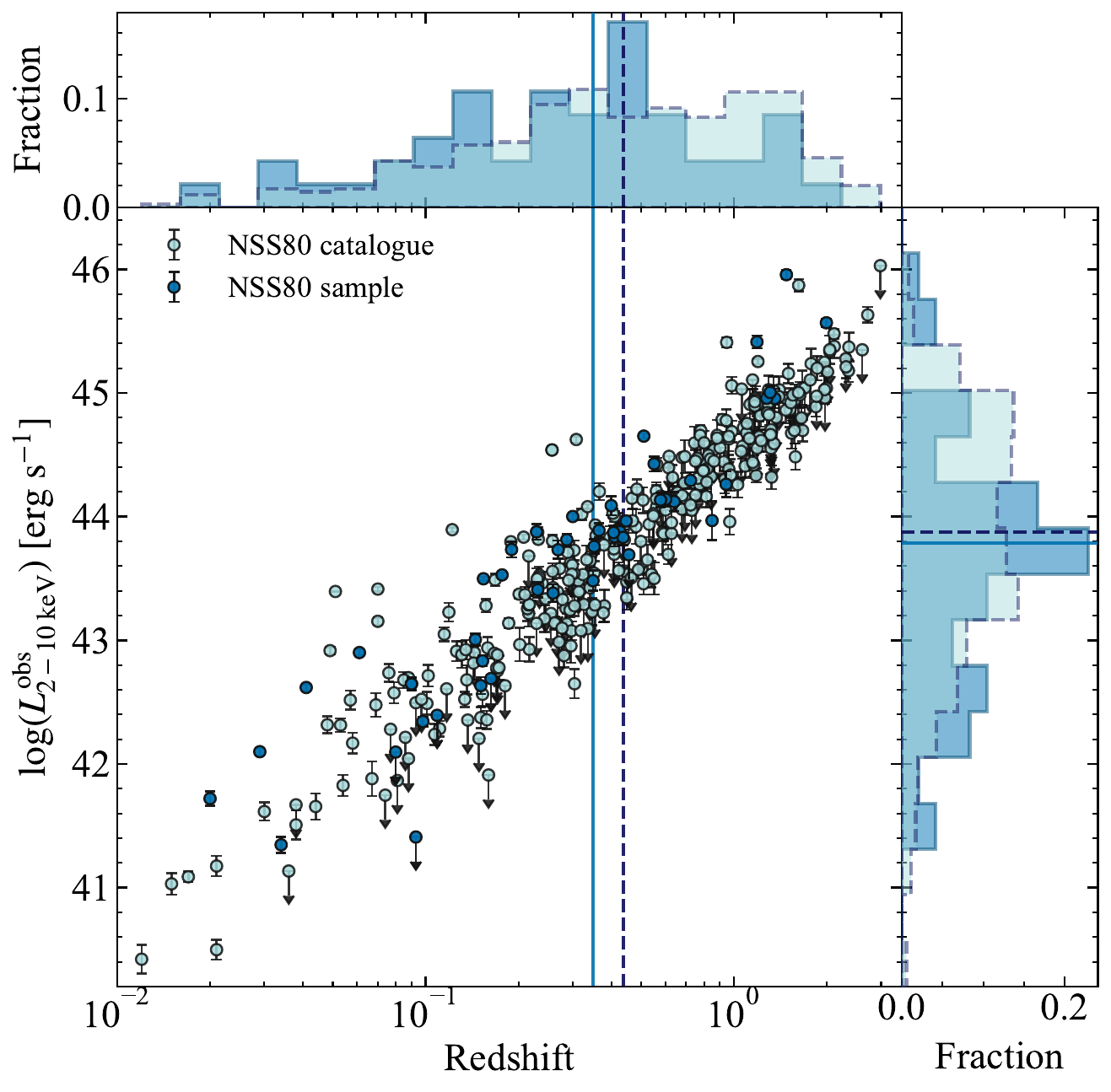}
    \caption{The distribution of estimated 2-10\,keV luminosity ($L_{2-10\mathrm{keV}}$) against redshift of the 350 valid NSS80 catalogue sources (light blue). The sources included within the NSS80 sample are highlighted in dark blue. The top panel plots the redshift distribution for the NSS80 catalogue (light blue dashed) compared to the NSS80 sample (blue), with the right panel showing the estimated 2-10\,keV luminosity derived from the NSS80 catalogue soft X-ray flux \citep{2024ApJS..273...20G}. The median redshift of the NSS80 catalogue and NSS80 sample are shown as vertical lines (dark blue dashed and solid blue respectively). The median X-ray luminosity is plotted as a horizontal line on the right panel. Scatter points with a downward arrow in the central panel correspond to upper bound sources, with no error on the soft X-ray flux. We note the consistent distributions between the NSS80 sample and the catalogue 2-10\,keV luminosity.}
    \label{fig:NSS80 catalogue luminosity redshift}
\end{figure}

To determine the CT fraction we compare the logarithmic luminosity ratio ($\log\left( {L}_{2-10\mathrm{keV}} / {L}_{12\mathrm{\mu m}}\right)$) of an NSS80 catalogue source to the CT AGN probability distribution using a similar methodology to section~\ref{sec: The CT probability} (see Fig.~\ref{fig:CT probability realisations}). We applied the following selection criteria to the NSS80 catalogue in order to fit the sources to the NSS80 X-ray-MIR diagnostic. We required a good-quality optical spectrum to determine redshift and therefore rest-frame X-ray and MIR luminosities. For the 12 $\mathrm{\mu m}$ luminosity the \textit{WISE} W3 magnitude and its uncertainty was required in addition to a measurement of either the W4 or W2 magnitude and its corresponding uncertainty for the K-correction. Blazars and Galactic sources were also removed from the candidate list at this stage. A large number of NSS80 catalogue sources do not have sufficient source counts for spectral fitting, so we instead used the deblended soft X-ray flux ($F_{3-8\mathrm{keV}}$) to estimate the 2-10\,keV luminosity. The uncertainty on the X-ray flux was not required such that the luminosity ratio of the catalogue would be an upper bound in the event of no soft X-ray flux uncertainty measurement. This choice allows for a lower bound on the CT fraction to be determined, rather than an upper bound. The selection left 350 AGN from the NSS80 catalogue, of which 103 sources have an upper bound on $L_{2-10\mathrm{keV}}$. The luminosity against redshift distribution of the NSS80 catalogue can be seen in Fig.\ref{fig:NSS80 catalogue luminosity redshift}.

We estimate the 2-10\,keV luminosity by assuming a simple power-law $\left(N\propto E^{-\Gamma}\right)$ with a photon index of $\Gamma = 1.8$ \citep{revisited_unified}. The photon index corresponds to the standard value used within the NSS80 \citep{2024ApJS..273...20G}, and was therefore used to be consistent with the approximations used within the catalogue, despite the standard value of $\Gamma = 1.9$ used in spectral fitting (section~\ref{sec: Spectral Modelling}). The difference between the photon index values however is largely negligible. To determine the accuracy of the luminosity approximation, the NSS80 sample of 47 sources were used to compare the observed spectral fitted luminosity to the approximate 2-10\,keV luminosity from the soft X-ray flux. The luminosity values were found to be within a factor of 2 for low obscuration sources ($\log(N_\mathrm{H}$/cm$^{-2})\lesssim23$), however for more heavily obscured sources, the approximation significantly overestimates the 2-10\,keV luminosity by a factor of $\sim5$ due to the power-law approximation. The overestimate was accounted for by applying a random multiplicative correction of 0.2 to the linear X-ray luminosity. The probability of the random correction was determined by the proportion of heavily obscured sources ($\log(N_\mathrm{H}$/cm$^{-2})> 23$) within the NSS80 sample sources, at $\sim35\%$. The effect of the luminosity correction can be seen in Fig.~\ref{fig:Obscuration Distribution} and ~\ref{fig:CT fraction distribution}.

As seen in Fig.~\ref{fig:NSS80 catalogue luminosity redshift}, the NSS80 catalogue sample used in this analysis has a redshift range of $z=0.012-2.983$ and median value of $\langle z \rangle=0.44$ which is comparable to the redshift of the X-ray spectral fitted NSS80 sample of $\langle z \rangle=0.35$. The approximate median 2-10\,keV luminosity of the NSS80 catalogue sample is comparable, but slightly more luminous than the NSS80 sample at $\langle L_{2-10\mathrm{keV}} \rangle = 7.5\times 10^{43}$ erg s $^{-1}$ and $\langle L_{2-10\mathrm{keV}} \rangle = 6.1\times 10^{43}$ erg s $^{-1}$. We further verify the consistency of the estimated 2-10\,keV luminosity distributions with an AD test statistic of $p=.13$. These X-ray luminosity differences are sufficiently small that the slight luminosity dependence on the X-ray-MIR diagnostic should not impact our results of the CT fraction ($f_{CT}$). 

\subsection{Analysis of the applied model}
\label{sec: Application to the NSS80}

Given the probability of a CT AGN is very low for sources with a high $L_X/L_{MIR}$ ratios (see Fig.~\ref{fig:CT probability realisations}), for this section we only include sources with $\log\left( {L}_{2-10\mathrm{keV}} / {L}_{12\mathrm{\mu m}}\right)< -0.062$\footnote{This limit corresponds to the predicted value of a source with column density $\log(N_\mathrm{H}$/cm$^{-2})=20$ using the NSS80 X-ray-MIR diagnostic.} in our application of the NSS80 X-ray-MIR diagnostic with the NSS80 catalogue. The following section uses the wider NSS80 catalogue (350 sources) to determine the CT fraction. Sources in the catalogue with a luminosity ratio greater than the threshold can be considered as likely unobscured. Application of this cut-off to the NSS80, reduces the number of sources to 196 with 51 upper bounds

\begin{figure}
	\includegraphics[width=\columnwidth]{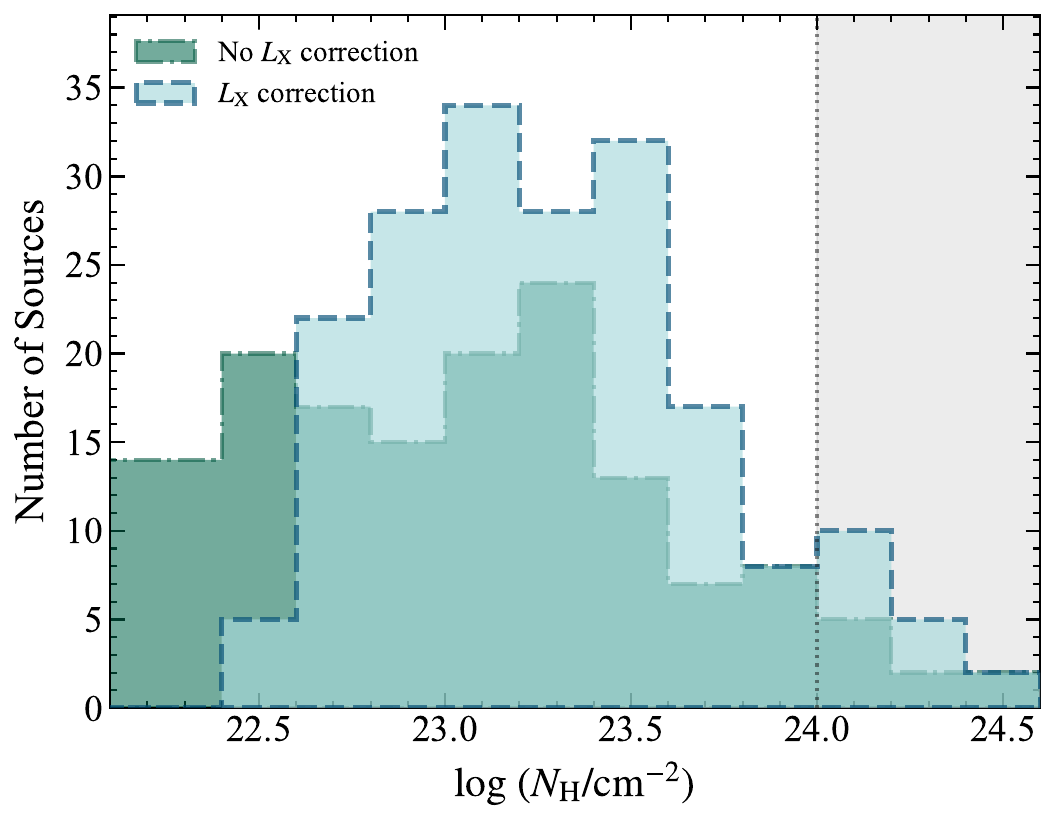}
    \caption{The column density ($\log(N_\mathrm{H}$/cm$^{-2})$) distribution of the NSS80 catalogue sample. The green dot-dashed histogram shows the distribution of NSS80 catalogue sources without the random correction applied to the X-ray luminosity, and the light blue dashed histogram represents the average distribution with the random multiplicative correction to the X-ray luminosity applied. The grey region denotes CT sources ($\log(N_\mathrm{H}$/cm$^{-2})\geq24$).}
    \label{fig:Obscuration Distribution}
\end{figure}

Fig.~\ref{fig:Obscuration Distribution} shows the predicted $\log(N_\mathrm{H}$/cm$^{-2})$ distribution of the 196 NSS80 catalogue sources with and without the X-ray luminosity correction. Both samples are predicted to have some CT sources ($\log(N_\mathrm{H}$/cm$^{-2})>24$) within the sample, with on average 9 likely CT sources without correction and 17 with the $L_X$ correction.

We used the NSS80 X-ray-MIR diagnostic on the NSS80 sample of 47 sources to compare the predicted column densities, to those determined by spectral fitting. We found 4/5 of the CT sources from spectral fitting, were also predicted to be CT by the NSS80 X-ray-MIR diagnostic. The exception was NuSTARJ194234-1011.9 predicted at $\log(N_\mathrm{H}$/cm$^{-2})=22.51^{+0.49}_{-0.46}$, and can be seen with large uncertainty to the right of the NSS80 sources in Fig.~\ref{fig:Project X-ray scatter}, but not outside the scatter seen with the \citetalias{pfeifle2022bass} and \citetalias{Asmus_diagnostic} samples.

The Band Ratio (BR)\footnote{The NSS80 \citep{2024ApJS..273...20G} uses the 3-8 keV and 8-24 keV count rates for the BR.} was used as a further comparison to the NSS80 X-ray-MIR diagnostic on the NSS80 catalogue. 
There were a total of 17 sources within the reduced catalogue sample with BR$>1.7$, which predicts the source is likely CT (in this case CT is defined as $N_\mathrm{H} > 1.5 \times 10^{24}$ cm$^{-2}$; \citealp{2024ApJS..273...20G}). Of the BR predicted CT sources, the X-ray-MIR diagnostic identified 13 as CT with the luminosity correction applied. All remaining sources were Compton thin within their uncertainty\footnote{Two of the BR flagged CT sources were Compton thick within uncertainty when considering the limit of this paper ($\log(N_\mathrm{H}$/cm$^{-2})>24$; NuSTARJ163126+2357.0 and NuSTARJ194234-1011.9).} with the lowest obscuration (NuSTARJ082303-0502.7) with $\log(N_\mathrm{H}$/cm$^{-2})=23.54^{+0.27}_{-0.27}$.  The X-ray-MIR diagnostic extends the CT prediction with an additional 3 CT sources within uncertainty that are not identified in the NSS80 with the BR (NuSTARJ010736-1732.3, NuSTARJ071422+3523.9 and NuSTARJ214239+4330.9), which will provide a basis for future research. The close agreement between the BR and the X-ray-MIR diagnostic on the predicted CT sources provides further validation of the X-ray-MIR diagnostic to identify potential CT AGN.

\subsection{Analysis of the Compton thick fraction}
\label{sec: CT fraction analysis}

\begin{figure}
	\includegraphics[width=\columnwidth]{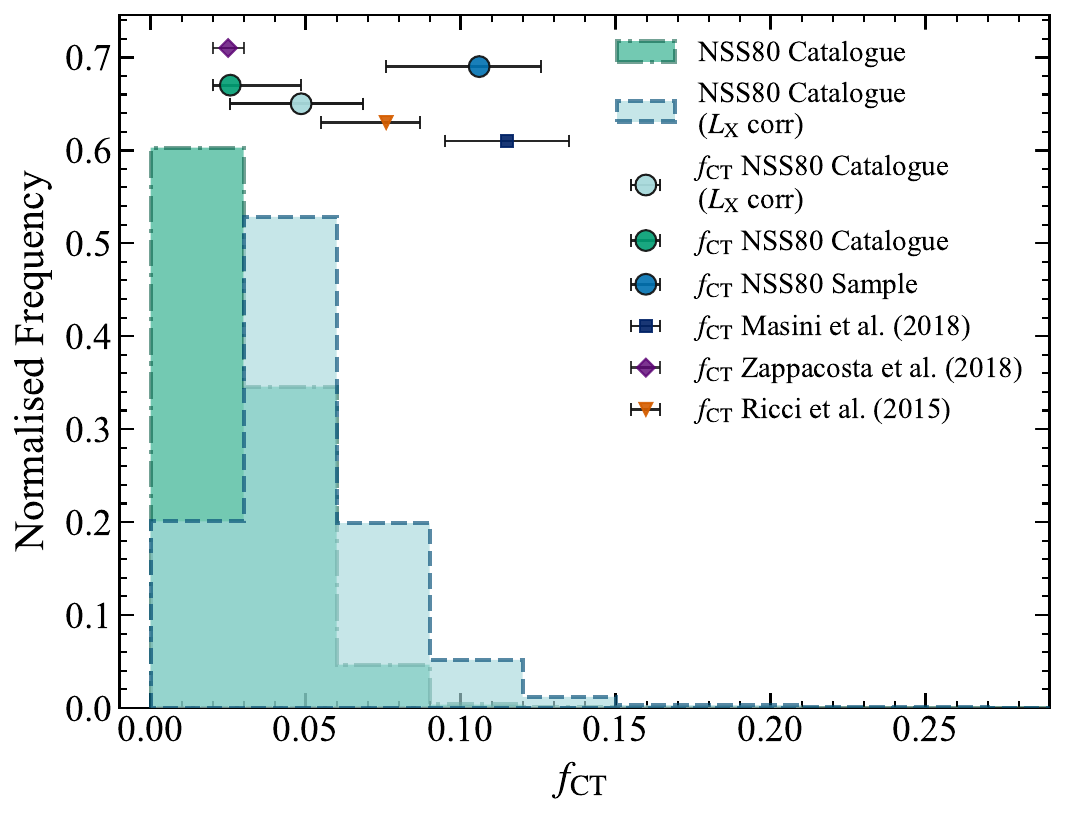}
    \caption{The fraction of CT AGN within the NSS80 catalogue inferred by applying the NSS80 X-ray-MIR diagnostic realisations using the same methods as section~\ref{sec: The CT probability}. The light blue dashed histogram shows the $f_{CT}$ distribution for the NSS80 catalogue sample, with the green dot-dashed histogram representing the distribution without the X-ray luminosity correction ($L_{\mathrm{X}}$ corr) applied. The error bars above represent the median value $f_{CT}$ with the 16 and 84 percentile values. The literature values of \citet{Ricci_2015} (orange filled triangle), \citet{2018NuSTAR_HardBand} (purple filled diamond) and \citet{Masini_2018} (dark blue filled square) of $f_{CT}$ are included for comparison.}
    \label{fig:CT fraction distribution}
\end{figure}

Following the predicted $\log(N_\mathrm{H}$/cm$^{-2})$ distribution of the reduced NSS80 catalogue, the CT fraction ($f_{\mathrm{CT}}$) was derived using a similar methodology to section~\ref{sec: The CT probability}. Within each MC realisation of the NSS80 $\log\left( {L}_{2-10\mathrm{keV}} / {L}_{12\mathrm{\mu m}}\right)$ diagnostic fit, the number of CT sources were counted. $f_{\mathrm{CT}}$ was then determined by that realisation as a fraction of the whole NSS80 catalogue sample (the 350 sources). The distribution of $f_{\mathrm{CT}}$ was then derived for all realisations, as seen in Fig.~\ref{fig:CT fraction distribution}.

The X-ray-MIR diagnostic fit for the NSS80 catalogue predicts a CT fraction of $f_{\mathrm{CT}}=4.9^{+2.0}_{-2.3}\%$ (see Fig.~\ref{fig:CT fraction distribution}) with the luminosity correction applied. The true value of $f_{\mathrm{CT}}$ may be higher, due to the catalogue sources with only an upper bound on 2-10\,keV luminosity available. A lower source $\log\left( {L}_{2-10\mathrm{keV}} / {L}_{12\mathrm{\mu m}}\right)$ would increase the statistical measure of $f_{\mathrm{CT}}$ (see Fig.~\ref{fig:CT probability realisations}).

It should be noted that the approximations used to derive ${L}_{2-10\mathrm{keV}}$ from the NSS80 soft X-ray flux have a large impact on the overall CT fraction. The uncorrected CT fraction was found to be lower at $f_{CT}=2.6^{+2.3}_{-0.6}\%$, compared to the catalogue sample with the luminosity correction, due to the overestimation of X-ray luminosity for heavily obscured sources. Not correcting the X-ray luminosity estimation results in larger average values of $\log\left( {L}_{2-10\mathrm{keV}} / {L}_{12\mathrm{\mu m}}\right)$, and therefore a lower probability of identifying an AGN as CT (as seen in Fig.\ref{fig:CT probability realisations}). 

The current measurements of the observed CT fraction vary depending on the luminosity of the sample, with a decrease in $f_{CT}$ for a higher luminosity sample \citep{Ricci_2015}. Moderately obscured and CT AGN generally have a lower observed X-ray luminosity than the unobscured AGN due to attenuation from the torus. Previous findings of \citet{Ricci_2015} using the \textit{Swift/BAT} 70-month catalogue predicted an observed CT fraction of $f_{CT}=7.6^{+1.1}_{-2.1}\%$ using X-ray spectral fitting. The sample used by \citet{Ricci_2015} is local with an average redshift of $z\approx0.055$. We find the $f_{CT}$ predicted by the X-ray-MIR diagnostic (blue circle in Fig.~\ref{fig:CT fraction distribution}), is in agreement with the value observed by \citet{Ricci_2015}. The \citet{2018NuSTAR_HardBand} sample of 63 sources with a median redshift of $\langle z \rangle = 0.58$ follows a very similar selection criteria (see section~\ref{sec: sample selection criteria}) reported $f_{CT}=2-3\%$ using spectral fitting. The CT fraction is lower but in agreement with the reported result of this paper using the NSS80 X-ray-MIR diagnostic, further validating the statistical measure of the CT fraction. 

We use the \citet{Masini_2018} sample with the \textit{NuSTAR} Ultra Deep Survey (UDS) field as a very high redshift comparison ($\langle z \rangle \sim 1.1$) to the NSS80 catalogue measurement. We found our measurement of $f_{CT}$ is in agreement within uncertainty with the CT fraction from \citet{Masini_2018} using only spectral analysis, at $f_{CT}=8.4\%\pm2.0\%$, but disagreement when including the Hardness Ratio (HR) measurements ($f_{CT}=11.5\%\pm2.0\%$; see Fig.~\ref{fig:CT fraction distribution}). The separation of $f_{CT}$ values are likely due to the UDS targeting faint X-ray flux AGN through long exposures \citep{Masini_2018}. The long exposures over a smaller area of $\sim$0.6 deg$^2$ would contribute to a higher sensitivity to obscured AGN, and a higher CT fraction than both the NSS80 and \citet{2018NuSTAR_HardBand} sample, despite the higher redshift of $\langle z \rangle = 1.1$ \citep{Masini_2018}.

We investigated the postulation that the lower $f_{CT}$ of the NSS80 catalogue relative to \citet{Masini_2018} is attributed to a low sensitivity in the NSS80 catalogue for CT sources with increasing redshift. We applied an arbitrary $z$ cut ($z<0.2$) to both the NSS80 sample and the reduced catalogue to remove high redshift sources. We found a significant increase in $f_{CT}$ for both the NSS80 sample with $f_{CT}=24^{+12}_{-9}\%$, and from the distribution of the realisations applied to the NSS80 catalogue with $f_{CT}= 18.3^{+6.1}_{-8.5}\%$. We therefore find that the initial value of $f_{CT}$ is possibly a result of the higher average redshift in the NSS80 catalogue, although this method needs a higher source count to draw a definitive conclusion. While the lower value of $f_{CT}$ could be attributed to the reduced NSS80 catalogue providing a lower bound, we ensured that the MIR was not underestimating $f_{CT}$ relative to other methods. The \citetalias{pfeifle2022bass} X-ray-MIR diagnostic was used with the same methodology as described on the \citetalias{pfeifle2022bass} sample, which predicted a value of $f_{CT}=11.0\%\pm1.1\%$, in good agreement with \citet{Masini_2018}, implying that the X-ray-MIR diagnostic is not underestimating $f_{CT}$.

\section{Conclusions}
\label{sec: Conclusions}
In this paper we have presented an analysis of the \citetalias{pfeifle2022bass} X-ray-MIR diagnostic for absorption, through an extension of the diagnostic at higher redshifts using the NSS80 catalogue \citep{2024ApJS..273...20G}. Our findings are summarised as follows:
\begin{enumerate}
    \item We determined $N_{\mathrm{H}}$ and $L_X$ constraints using spectral fitting of two physically motivated AGN models on 47 NSS80 sources with \textit{NuSTAR} and \textit{XMM-Newton} observations.
    \item The X-ray-MIR diagnostic applied to the NSS80 is consistent with the shape of the model from \citetalias{pfeifle2022bass} but with an initial offset in $L_X/L_{MIR}$ ratio. The offset can be attributed to a higher median X-ray luminosity within the NSS80 sample, implying that using the X-ray-MIR diagnostic requires an initial understanding of the X-ray luminosity distribution within a sample.
    \item We derived the probability of a source to be Compton Thick given $\log\left( {L}_{2-10\mathrm{keV}} / {L}_{12\mathrm{\mu m}}\right)$. Using the NSS80 X-ray-MIR diagnostic we predicted a source to be CT with 90$\%$ confidence at $\log\left( {L}_{2-10\mathrm{keV}} / {L}_{12\mathrm{\mu m}}\right)<-1.83$, allowing for a simpler method of identifying CT AGN.
    \item We showed that the NSS80 X-ray-MIR diagnostic is consistent with spectral fitting, predicting 4/5 CT sources within the NSS80 sample. We also found consistent measurements to the band ratio, with 13/17 sources predicted to be CT. We further identify three new potential CT sources within the NSS80 catalogue that were not identified by the band ratio (NuSTARJ010736-1732.3, NuSTARJ071422+3523.9 and NuSTARJ214239+4330.9).
    \item We found the CT fraction within the NSS80 catalogue to be $f_{\mathrm{CT}}=4.9^{+2.0}_{-2.3}\%$, in agreement with \citet{Ricci_2015} and \citet{2018NuSTAR_HardBand}, but lower than the high redshift \textit{NuSTAR} sample from \citet{Masini_2018}.
\end{enumerate}

We conclude that the X-ray-MIR diagnostic for absorption can be extended to high $z$ AGN, following the close fit of the NSS80 X-ray-MIR diagnostic relative to the low-$z$ \citetalias{pfeifle2022bass} sample. This work provides a method for predicting the column density of a source and allow CT AGN to be identified, even with too few counts for spectral fitting. The X-ray-MIR diagnostic would provide an effective method of identifying CT AGN in soft X-ray surveys, with a high number of sources, such as \textit{Chandra} and \textit{XMM-Newton}. In future, a study with a larger sample of high $z$ AGN with reliable measurements of $N_{\mathrm{H}}$ from direct X-ray spectral fitting would allow us to better understand the luminosity dependence in the X-ray-MIR diagnostic.

\section*{Acknowledgements}

We would like to thank the anonymous referee for their insightful and constructive comments that helped improve this paper. We would also like to thank Ryan W. Pfeifle \citepalias{pfeifle2022bass} for his contributions to this research in kindly providing the sample data, which was invaluable to our work. We thank Chris Done for her invaluable comments on this work.

DMA and CLG acknowledge support from the Science and Technology Facilities Council (grant codes: ST/T000244/1 and ST/X001075/1).

This research has made use of data from the NuSTAR mission, a project led by the California Institute of Technology, managed by the Jet Propulsion Laboratory, and funded by the National Aeronautics and Space Administration. Data analysis was performed using the NuSTAR Data Analysis Software (NuSTARDAS), jointly developed by the ASI Science Data Center (SSDC, Italy) and the California Institute of Technology (USA). This research has made use of data and/or software provided by the High Energy Astrophysics Science Archive Research Center (HEASARC), which is a service of the Astrophysics Science Division at NASA/GSFC. This research is based on observations obtained with XMM-Newton, an ESA science mission with instruments and contributions directly funded by ESA Member States and NASA. This publication makes use of data products from the Wide-field Infrared Survey Explorer, which is a joint project of the University of California, Los Angeles, and the Jet Propulsion Laboratory/California Institute of Technology, funded by the National Aeronautics and Space Administration.

This paper makes use of the following software and libraries throughout the research: AstroPy \citep{astropy:2013, astropy:2018, astropy:2022}, NumPy \citep{harris2020array}, Matplotlib \citep{Hunter:2007}, SciPy \citep{VirtanenSCIPY} and Topcat \citep{Topcat}.

\section*{Data Availability}

The data underlying this article were accessed from public archives, and the derived data are available in the article and its online supplementary material and/ or will be shared on reasonable request to the corresponding author. The data from \citet{pfeifle2022bass} were provided by Dr R.W. Pfeifle. Data will be shared on request to the corresponding author with permission of Dr R.W. Pfeifle.

\nocite{Ricci_2017_b}
\bibliographystyle{mnras}
\bibliography{Report_bib}

\appendix

\section{NuSTAR sample source information}
\label{Appendix: Source info}

Table \ref{tab:Observation} shows the observation data used in the spectral fitting of the NSS80 sample.
\onecolumn 
\begin{landscape}
            
    \renewcommand{\arraystretch}{1.5}
    \setlength{\LTcapwidth}{\textwidth} 
    \begin{longtable}{@{\extracolsep{\fill}}lcccccccc}
        \caption{Observation details of the NSS80 sample.} \label{tab:Observation} \\ 
        \hline
          Source & RA & DE & NuSTAR Obs ID & NuSTAR Date& XMM-Newton Obs ID & XMM-Newton Date & $z$ & Net Counts \\
          (1) & (2) & (3) & (4) & (5) & (6) & (7) & (8) & (9) \\
        \hline
        \endfirsthead
    
        \caption[]{Observation details of the NSS80 sample (Continued).} \\
        \hline
          Source & RA & DE & NuSTAR Obs ID & NuSTAR Date& XMM-Newton Obs ID & XMM-Newton Date & $z$ & Net Counts \\
          (1) & (2) & (3) & (4) & (5) & (6) & (7) & (8) & (9) \\
        \hline
        \endhead
    
        \hline
        \multicolumn{9}{r}{Continued} \\
        \endfoot
    
        \hline
        \multicolumn{9}{l}{%
             \begin{minipage}{\textwidth}
              \vspace{2mm}
              \textbf{Notes:} (1) The \textit{NuSTAR} source name within the NSS80 catalogue. 
              (2 and 3) The Right Ascension (RA) and Declination (DE) corresponding to the optical counterpart coordinates used for matching in HEASARC's Xamin. 
              (4) The \textit{NuSTAR} observation ID used. 
              (5) The date of the selected \textit{NuSTAR} observation (YYYY-MM-DD).
              (6) The selected \textit{XMM-Newton} observation ID used. 
              (7) The date of the selected \textit{XMM-Newton} observation (YYYY-MM-DD).
              (8) The source redshift from the NSS80 catalogue. 
              (9) The (3-24\,keV) net deblended counts of the source from the NSS80 catalogue.
             \end{minipage}%
        } \\
        \endlastfoot
        
        NuSTARJ001443+8131.9 & $3.680$ & $81.537$ & $60001098004$ & 2015-01-23 & $0112620201$ & 2001-08-23 & $0.365$ & $149$ \\
        NuSTARJ012618+2559.2 & $21.581$ & $25.988$ & $60201047002$ &2017-01-15 & $0790820101$ & 2017-01-15 & $0.41$ & $110$ \\
        NuSTARJ012633+2556.2 & $21.641$ & $25.936$ & $60201047002$ &2017-01-15 & $0790820101$ &2017-01-15 & $0.427$ & $161$ \\
        NuSTARJ021435-0042.7 & $33.645$ & $-0.712$ & $90201043002$ &2016-12-02 & $0201020201$ &2004-07-04 & $0.437$ & $205$ \\
        NuSTARJ022742+3331.5 & $36.929$ & $33.525$ & $30201003002$ &2017-01-18 & $0784510301$ &2017-01-18 & $0.09$ & $229$ \\
        NuSTARJ023229+2023.7 & $38.120$ & $20.397$ & $60002047004$ &2013-10-05 & $0604210201$ &2009-08-21 & $0.029$ & $467$ \\
        NuSTARJ024144+0512.3 & $40.434$ & $5.206$ & $60363003002$ &2017-07-13 & $0802200401$ &2017-07-13 & $0.07$ & $648$ \\
        NuSTARJ031015-7651.5 & $47.566$ & $-76.859$ & $60160145002$ &2016-03-19 & $0122520201$ &2000-03-31 & $1.187$ & $161$ \\
        NuSTARJ032459-0256.2 & $51.250$ & $-2.937$ & $60061036002$ &2012-10-25 & $0405240201$ &2006-08-06 & $0.02$ & $121$ \\
        NuSTARJ042613-5714.9 & $66.554$ & $-57.253$ & $60101039002$ &2015-06-03 & $0148000601$ &2003-09-16 & $0.098$ & $527$ \\
        NuSTARJ045348+0408.6 & $73.451$ & $4.142$ & $60001158002$ & 2014-11-10 &$0307000401$ &2005-08-30 & $1.357$ & $160$ \\
        NuSTARJ051937-3233.3 & $79.907$ & $-32.559$ & $60201046002$ & 2016-09-24 &$0790810101$ &2016-09-24 & $0.64$ & $143$ \\
        NuSTARJ052531-4557.8 & $81.381$ & $-45.965$ & $60101014002$ &2015-08-21 & $0762920501$ &2015-08-19 & $1.479$ & $313$ \\
        NuSTARJ090053+3856.3 & $135.224$ & $38.938$ & $60001088002$ &2013-05-23 & $0200630101$ &2004-04-13 & $0.229$ & $147$ \\
        NuSTARJ094910+0022.9 & $147.2942$ & $0.3802$ & $60201052002$ & 2016-11-04 &$0673730101$ &2011-05-28 & $0.093$ & $117$ \\
        NuSTARJ102345+0044.1 & $155.939$ & $0.735$ & $30001027006$ &2013-10-19 & $0560180801$ & 2008-11-26 &$0.3$ & $951$ \\
        NuSTARJ103135-4206.0 & $157.900$ & $-42.103$ & $60468002002$ &2018-05-27 & $0830500201$ &2018-05-27 & $0.061$ & $984$ \\
        NuSTARJ103456+3939.6 & $158.735$ & $39.6614$ & $60401004002$ &2018-10-30 & $0506440101$ &2007-05-31 & $0.151$ & $411$ \\
        NuSTARJ112745-2911.0 & $171.939$ & $-29.181$ & $60101012002$ &2015-12-20 & $0762920301$ &2015-06-07 & $0.19$ & $129$ \\
        NuSTARJ115658+5508.2 & $179.241$ & $55.139$ & $60375001002$ &2017-12-06 & $0204651201$ &2004-06-15 & $0.08$ & $116$ \\
        NuSTARJ115912+4232.6 & $179.801$ & $42.543$ & $60061217004$ &2013-02-04 & $0744040301$ &2014-11-28 & $0.177$ & $997$ \\
        NuSTARJ120348+4428.2 & $180.949$ & $44.470$ & $60401009002$ &2018-11-04 & $0109141401$ & 2001-05-16 &$1.994$ & $547$ \\
        NuSTARJ121355+1404.4 & $183.484$ & $14.075$ & $60001100002$ & 2014-02-18 &$0745110501$ &2014-06-25 & $0.154$ & $1856$ \\
        NuSTARJ121426+1403.1 & $183.611$ & $14.050$ & $60001100002$ & 2014-02-18 &$0502050101$ &2007-12-21 & $1.277$ & $652$ \\
        NuSTARJ124946+2629.0 & $192.443$ & $26.488$ & $60001030002$ & 2014-12-04 &$0143150201$ &2003-06-18 & $0.407$ & $209$ \\
        NuSTARJ132231-1647.4 & $200.629$ & $-16.790$ & $60101020002$ &2016-01-17 & $0206580101$ & 2005-01-24 &$0.349$ & $111$ \\
        NuSTARJ133401-3403.6 & $203.506$ & $-34.063$ & $60261002002$ & 2016-01-20 &$0307000901$ &2006-01-10 & $0.145$ & $235$ \\
        NuSTARJ134934-3025.5 & $207.388$ & $-30.427$ & $60001045002$ &2012-08-12 & $0101040401$ &2001-01-31 & $0.163$ & $122$ \\
        NuSTARJ140536+2551.7 & $211.403$ & $25.861$ & $60401016002$ &2018-12-14 & $0400200101$ &2006-12-21 & $0.943$ & $107$ \\
        NuSTARJ141809+2500.7 & $214.537$ & $25.012$ & $60002044006$ &2013-09-10 & $0089960301$ &2001-07-09 & $0.723$ & $174$ \\
        NuSTARJ141914+2104.2 & $214.805$ & $21.071$ & $50101003002$ &2016-01-01 & $0782380101$ &2016-06-28 & $0.109$ & $123$ \\
        NuSTARJ143026+4159.9 & $217.607$ & $41.999$ & $60001103002$ &2014-07-14 & $0111260701$ & 2003-01-17 &$0.352$ & $148$ \\
        NuSTARJ150646+0346.2 & $226.691$ & $3.773$ & $60301023002$ &2018-01-20 & $0795670101$ &2018-01-20 & $0.034$ & $445$ \\
        NuSTARJ151653+5617.7 & $229.224$ & $56.296$ & $30002039005$ &2013-11-12 & $0804090701$ &2017-07-12 & $1.31$ & $286$ \\
        NuSTARJ152722+3600.2 & $231.841$ & $36.005$ & $60374009002$ &2018-01-17 & $0081341101$ & 2002-02-22 &$0.269$ & $102$ \\
        NuSTARJ153640+5435.1 & $234.174$ & $54.585$ & $60201023004$ &2016-09-14 & $0790590201$ &2016-09-14 & $0.447$ & $470$ \\
        NuSTARJ161445+4659.3 & $243.693$ & $46.990$ & $60301021002$ &2017-08-25 & $0803910101$ &2017-08-05 & $0.153$ & $373$ \\
        NuSTARJ165105-0129.4 & $252.774$ & $-1.491$ & $60001145002$ &2015-03-03 & $0743010301$ &2015-03-04 & $0.041$ & $551$ \\
        NuSTARJ183442+3237.9 & $278.680$ & $32.633$ & $60202015004$ &2016-09-11 & $0506120101$ &2008-04-28 & $0.51$ & $1085$ \\
        NuSTARJ192215-5844.9 & $290.563$ & $-58.752$ & $60201042002$ &2016-07-15 & $0503750501$ &2007-10-12 & $0.231$ & $166$ \\
        NuSTARJ193246-7233.8 & $293.202$ & $-72.565$ & $60002029002$ &2013-05-26 & $0722610401$ &2013-09-02 & $0.287$ & $211$ \\
        NuSTARJ194234-1011.9 & $295.639$ & $-10.197$ & $60201028002$ & 2016-07-04 &$0550451801$ &2009-04-22 & $0.849$ & $131$ \\
        NuSTARJ200755-4439.7 & $301.980$ & $-44.662$ & $60201045002$ &2016-10-23 & $0790630101$ &2016-05-05 & $0.456$ & $116$ \\
        NuSTARJ202226+6141.3 & $305.610$ & $61.692$ & $60401023002$ &2018-07-07 & $0784610301$ &2016-05-25 & $0.259$ & $106$ \\
        NuSTARJ204020-0056.1 & $310.087$ & $-0.935$ & $30001120004$ &2012-09-05 & $0111180201$ &2001-11-07 & $0.601$ & $252$ \\
        NuSTARJ211517+0608.7 & $318.819$ & $6.145$ & $60001159002$ &2014-12-23 & $0500500701$ &2007-05-19 & $0.4$ & $112$ \\
        NuSTARJ213900-2640.4 & $324.749$ & $-26.676$ & $60362025002$ &2017-05-09 & $0805150801$ &2017-05-09 & $0.551$ & $136$ \\
        NuSTARJ221133+1841.5 & $332.890$ & $18.693$ & $60301015002$ & 2017-11-23 &$0801893301$ &2017-11-30 & $0.578$ & $182$ \\
    
    \end{longtable}

\end{landscape}
\twocolumn 

\section{Source variability}
\label{sec: Source Variability}

In addition to the checks on the model fits (see section~\ref{sec: Source checks}), we assessed the variability of the NSS80 sample, using the sample's soft X-ray \textit{XMM-Newton} flux against the \textit{Swift-XRT} 2SXPS catalogue \citep{2SXPS_catalogue}. Only one source, NuSTARJ024144+0512.3, was found to have extreme variability between \textit{XMM-Newton} and \textit{Swift-XRT}, with very limited counts in 2SXPS catalogue and a S/N of 1.9. Spectral fitting modelled the source as Compton thin using both the Power-law and Borus model (see Table~\ref{tab:Results}) with a cstat of 318/314 (for Borus) suggesting a good fit. It was found that removal of NuSTARJ024144+0512.3 did not significantly change our results in fitting the X-ray-MIR diagnostic (see section~\ref{sec: mir results}), so as a conservative choice we decided to remove the source from the NSS80 sample due to the significant variability between \textit{XMM-Newton} and \textit{Swift-XRT}.

\section{Analysis on X-ray-MIR diagnostic fitting}

\subsection{K-correction of \textit{WISE} magnitudes}
\label{sec:K-correction}

The \textit{WISE} magnitudes for each NSS80 source within the sample are K-corrected in the MIR to account for the effect of redshift. An approximation method was implemented by assuming a simple power-law  \citep[$\log(F_{\nu})\propto-\alpha \log(\nu)$;][]{intro_AGN}. The rest frame flux was calculated using the ratio of the fractional difference between the rest frequency and the frequency separation of the WISE bands. If the \textit{WISE} W4 magnitude or its error is not provided, then the 12\,$\mathrm{\mu m}$ flux is instead extrapolated using the W2 band magnitude. The uncertainty is carried forward through the fractional error in the flux of both \textit{WISE} bands used. 

The K-correction is small at low redshifts ($z<0.1$), and was therefore not applied to the \textit{Swift/BAT} samples, however, the extended range of the NSS80 sample up to $z=1.994$ requires the redshift correction.

\subsection{The inverse MC fitting method}
\label{app: inverse}

\begin{figure}
	\includegraphics[width=\columnwidth]{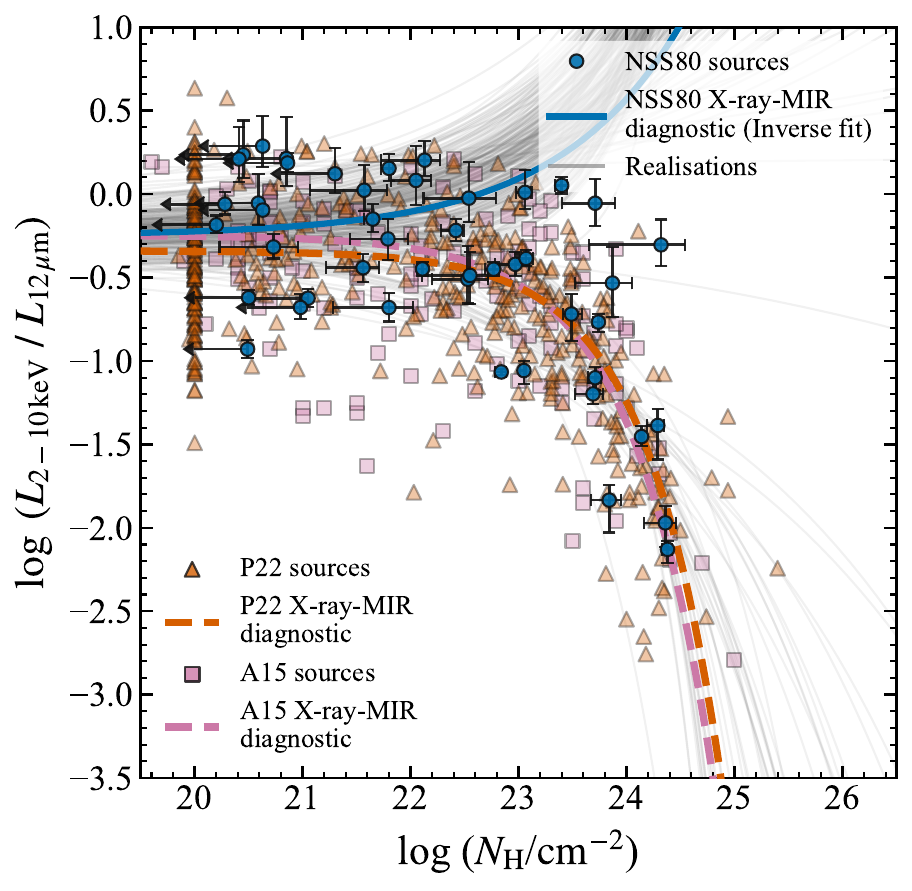}
    \caption{The distribution of the luminosity ratio ($\log\left( {L}_{2-10\mathrm{keV}} / {L}_{12\mathrm{\mu m}}\right)$) against column density ($\log(N_\mathrm{H}$/cm$^{-2})$) of the NSS80 sample (blue circles) compared to the \citetalias{pfeifle2022bass} (orange triangles) and \citetalias{Asmus_diagnostic} (pink squares) samples. The NSS80 X-ray-MIR diagnostic fit using the inverse of equation~\ref{eqn: default model} (blue line) is plotted with the \citetalias{Asmus_diagnostic} (pink dashed line) and \citetalias{pfeifle2022bass} (orange dashed line) X-ray-MIR diagnostics for comparison. A sample of 300 realisations (grey) are included from the NSS80 MC fit. The NSS80 X-ray luminosity was calculated using the XSPEC Lumin function, with \citetalias{Asmus_diagnostic} and \citetalias{pfeifle2022bass} luminosities measurements provided within the respective catalogues. Sources with an upper bound in column density are plotted using an arrow with the data point set at 1$\sigma$ uncertainty. Note the two distinct NSS80 X-ray-MIR curve trends from the realisations, showing a poor convergence from the inverse of equation~\ref{eqn: default model}.}
    \label{fig:NSS80 Sample Inverse fit}
\end{figure}

As discussed in section~\ref{sec: Sample fitting methods} the MC fitting with the inverse of equation~\ref{eqn: default model} does not converge to a set curve shape, with the distribution of ODR fits (the realisations) heavily dependant on the initial input parameters into the ODR fitting. Fig.~\ref{fig:NSS80 Sample Inverse fit} shows an example of the best fit model using the inverse of equation~\ref{eqn: default model}. There are two separate curve trends seen in the realisations, with  the secondary curve with the fewer realisations having a similar shape to the NSS80 X-ray-MIR diagnostic in Fig.~\ref{fig:Project X-ray scatter}. The primary shape of the realisations show increasing $L_X/L_{MIR}$ with column density.

The upward trend on the inverse NSS80 X-ray-MIR diagnostic can be ignored as this corresponds to the $\log\left( {L}_{2-10\mathrm{keV}} / {L}_{12\mathrm{\mu m}}\right)$ ratio range primarily for unobscured sources. The poor convergence of the inverse X-ray-MIR diagnostic is in agreement with the findings from the fitting procedure of \citetalias{pfeifle2022bass}. This work therefore used equation \ref{eqn: default model} for the MC model fitting, and then inverted the equation to predict the column density of catalogue sources in section~\ref{sec: Application to the NSS80}.

\subsection{The {\texorpdfstring{$L_{22\,\mu\mathrm{m}} / L_{4.6\,\mu\mathrm{m}}$}{L 22um / L 4.6um}} X-ray-MIR diagnostic}

\begin{figure}
	\includegraphics[width=\columnwidth]{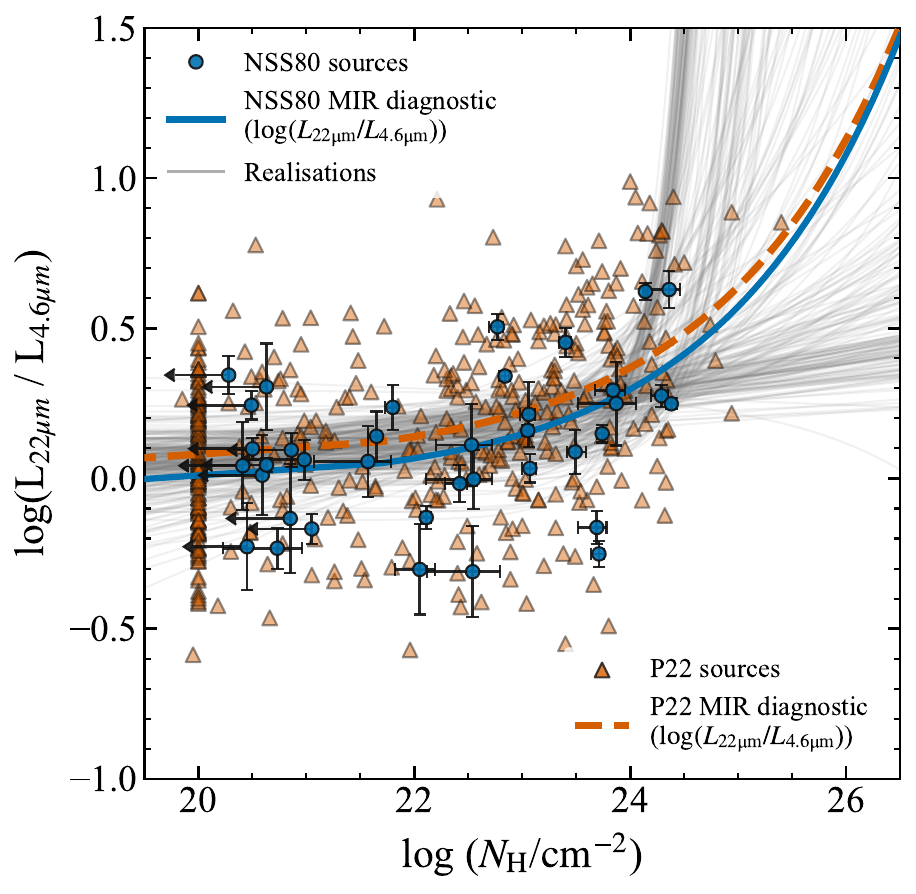}
    \caption{The distribution of the MIR luminosity ratio ($\log\left( L_{22\mathrm{\mu m}} / L_{4.6\mathrm{\mu m}}\right)$) against column density ($\log(N_\mathrm{H}$/cm$^{-2})$) of the NSS80 sample (blue circles) and the \citetalias{pfeifle2022bass} sample (orange triangles). 38 out of the 47 NSS80 sources are included here due to lack of constrained \textit{WISE} $W4$ measurement. The fully MIR diagnostic for the NSS80 is shown (blue line) with the best fit model from \citetalias{pfeifle2022bass} (orange dashed line). A sample of 300 realisations from the NSS80 MIR diagnostic is shown (grey) to illustrate the uncertainty on the fit. The NSS80 MIR luminosities were calculated from the \textit{WISE} MIR magnitudes provided within the NSS80 catalogue \citep{2024ApJS..273...20G}, and \citetalias{pfeifle2022bass} luminosities were provided in the respective catalogue. Sources with an upper bound in column density are plotted using an arrow with the data point set at 1$\sigma$ uncertainty. The large uncertainty on the NSS80 MIR diagnostic fit can be seen with a significant range in realisations at high obscuration implying an unconstrained MC fit.}
    \label{fig:Project MIR scatter}
\end{figure}

We assess the alternative $\log\left( L_{22\mathrm{\mu m}} / L_{4.6\mathrm{\mu m}}\right)$ diagnostic from \citetalias{pfeifle2022bass} using the NSS80 sample. The $\log\left( L_{22\mathrm{\mu m}} / L_{4.6\mathrm{\mu m}}\right)$ would allow for the prediction and identification of CT sources without the need for X-ray emission detection, which decreases for more obscured sources. As the alternative diagnostic method requires \textit{WISE} W4 and W2 bands for the corresponding MIR luminosity, we require all \textit{WISE} bands and their error in order to K-correct (which are interpolated as in section~\ref{sec:K-correction}). This reduced the NSS80 sample down to 38 sources. The methods of fitting the NSS80 sample are identical to those described in section~\ref{sec: Sample fitting methods} using the form of equation~\ref{eqn: default model}. The results of this fit against \citetalias{pfeifle2022bass} can be seen in Fig.~\ref{fig:Project MIR scatter}.

The large scatter within the $\log\left( L_{22\mathrm{\mu m}} / L_{4.6\mathrm{\mu m}}\right)$ X-ray-MIR diagnostic suggests that the relation is not as strong as $\log\left( {L}_{2-10\mathrm{keV}} / {L}_{12\mathrm{\mu m}}\right)$, likely due to contamination from the host galaxy significantly impacting the observed \textit{WISE} magnitudes. The overall best fit does appear in Fig.~\ref{fig:Project MIR scatter}, although it is poorly constrained, with two distinct branching paths in the realisations. We suggest that this method of identifying CT AGN is significantly less reliable compared to the $\log\left( {L}_{2-10\mathrm{keV}} / {L}_{12\mathrm{\mu m}}\right)$ X-ray-MIR diagnostic, hence we did not include this in the main body of our work.

\section{Analysis of the Luminosity Ratio}
\label{Appendix: Analysis of the Luminosity Ratio}

\begin{figure*}
    \includegraphics[width=0.9\textwidth]{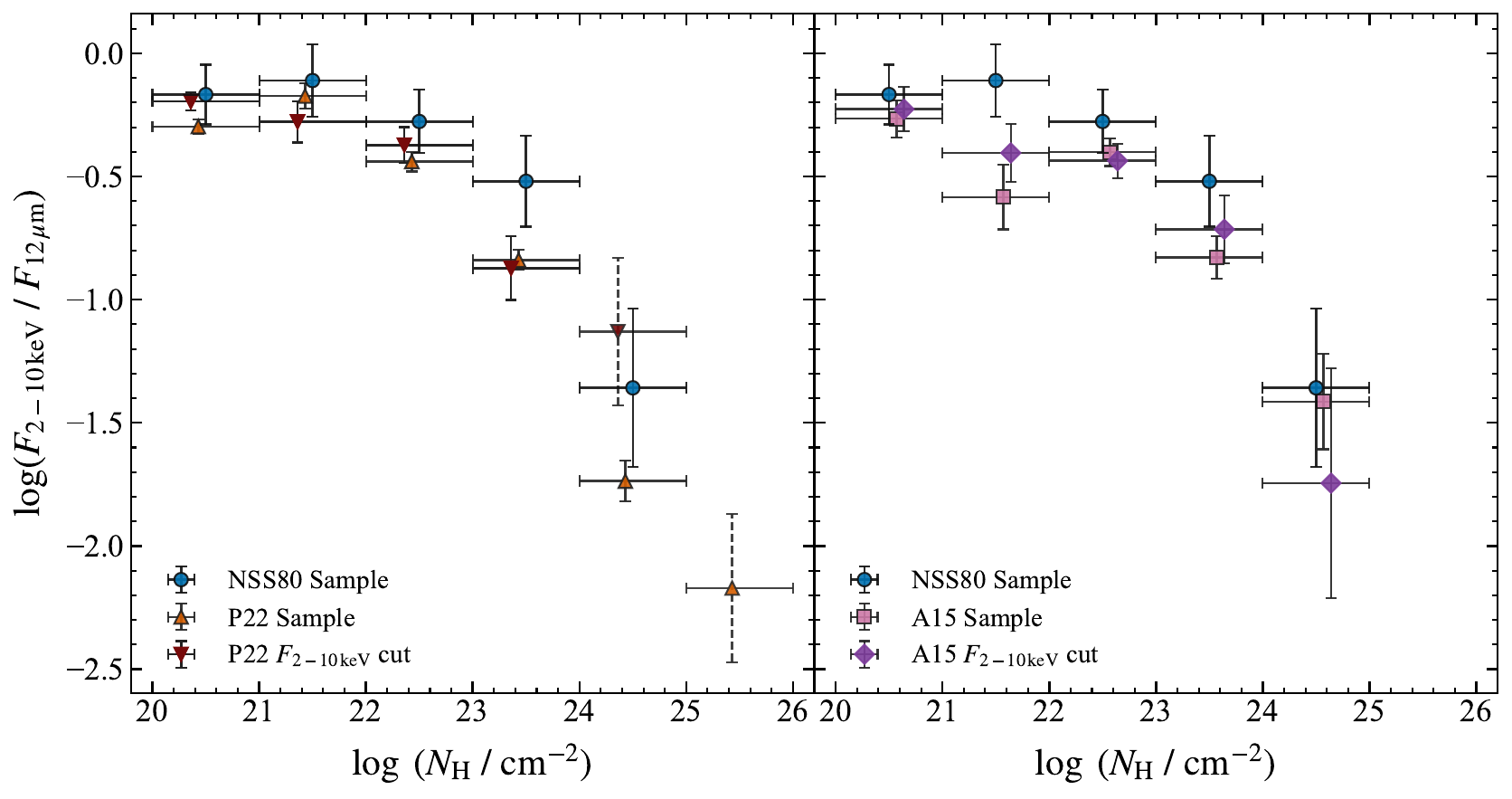}
    \caption{The distribution of $\log\left( F_{2-10\mathrm{keV}} / F_{12\mathrm{\mu m}}\right)$ plotted as a function of column density ($\log(N_\mathrm{H}$/cm$^{-2})$) with a bin width of 1. The NSS80 sample (blue circles) is compared to (left) the \citetalias{pfeifle2022bass} sample of \textit{Swift/BAT} AGN (orange triangles) and (right) the \citetalias{Asmus_diagnostic} sample of local AGN (pink squares). We compare the samples with a flux cut of $F_{2-10\mathrm{keV}}\geq6\times10^{-12}$\,erg\,s$^{-1}$\,cm $^{-2}$ applied to the \citetalias{pfeifle2022bass} (red inverted triangles) and \citetalias{Asmus_diagnostic} (purple diamonds) to determine the effect of a higher average X-ray flux on the sample distributions. The uncertainty is provided as the standard error of each respective bin except the bins with vertical dashed error bars, which include only one source. For these bins we instead use the uncertainty on the flux ratio for the given source. All bins for \citetalias{pfeifle2022bass} and \citetalias{Asmus_diagnostic} have been offset for clarity. Note for the NSS80 sample the X-ray flux was calculated using the XSPEC Flux function, with the \citetalias{Asmus_diagnostic} sample calculated from the provided luminosity, and \citetalias{pfeifle2022bass} from the provided flux measurements within the respective catalogues. We observe the flux cut subsamples follow closely to their full samples, suggesting that the separation between the NSS80 and the other samples is not dependant on the X-ray flux.}
    \label{fig:Appendix flux bin plot}

\end{figure*}

In addition to the X-ray luminosity cuts we tested in section~\ref{sec: mir results}, we applied X-ray flux cuts to the \citetalias{Asmus_diagnostic} and \citetalias{pfeifle2022bass} samples. The flux cut arbitrarily removed low X-ray flux sources with $F_{2-10\mathrm{keV}}\leq6\times10^{-12}$\,erg\,s$^{-1}$\,cm $^{-2}$ from both samples. Fig.~\ref{fig:Appendix flux bin plot} shows the distribution for the NSS80 sample with the \citetalias{pfeifle2022bass} and \citetalias{Asmus_diagnostic} samples with their flux cuts applied. It can be seen that the trends of the samples so not vary significantly with a flux cut, implying the difference in X-ray flux between samples does not result in the offset in X-ray-MIR diagnostic fit shown in Fig.~\ref{fig:Project X-ray scatter} and Table~\ref{tab:Parameters}. From this analysis, we discussed the influence of the luminosity cuts in section~\ref{sec: mir results}.

\begin{figure*}
    \includegraphics[width=\textwidth]{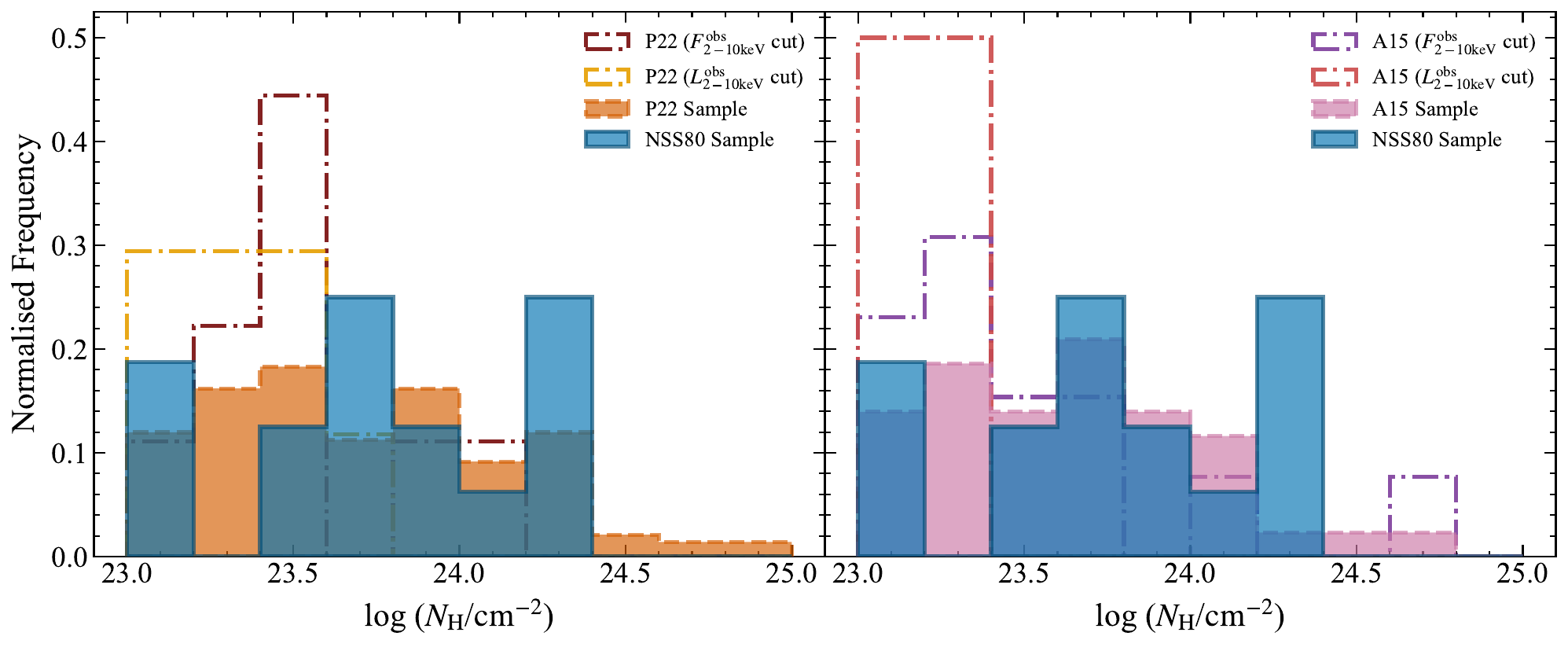}
    \caption{The column density distribution of the NSS80 sample (blue), \citetalias{pfeifle2022bass} sample (orange dashed; left) and \citetalias{Asmus_diagnostic} sample (pink dashed; right) in the range of $23<\log(N_\mathrm{H}/\mathrm{cm}^{-2})<25$. We plot the \citetalias{pfeifle2022bass} and \citetalias{Asmus_diagnostic} samples with a lower bound X-ray luminosity cut (yellow and red dot-dashed respectively) as well as a flux cut (dark red and purple dot-dashed respectively) to determine the effect of the X-ray cuts on the samples. The NSS80 sample column density is determined from X-ray spectral fitting, with the \citetalias{pfeifle2022bass} and \citetalias{Asmus_diagnostic} measurements provided within the catalogues.}
    \label{fig:Appendix column density}

\end{figure*}

We further investigated the distribution of $N_{H}$ from Compton thin and above sources between the NSS80 and \citetalias{pfeifle2022bass} samples. We also compared the distribution with \citetalias{Asmus_diagnostic}, however we emphasise that the comparison is not as informative as the \citetalias{pfeifle2022bass} sample due to the differing sample selection criteria and the use of the nuclear MIR luminosity. The distribution of the samples can be seen in Fig.~\ref{fig:Appendix column density}.

We performed Anderson-Darling (AD) tests on the samples with the luminosity and flux cuts against the NSS80 sample for the distributions shown in Fig.~\ref{fig:Appendix column density}. We find that the NSS80 sample is statistically similar to the full \citetalias{pfeifle2022bass} sample ($p>.25$) and the flux cut sample ($p=.16$) in $N_{\mathrm{H}}$ to the 5$\%$ significance level. The luminosity cut sample however is statistically different ($p=.002$), implying a different distribution. The luminosity cuts of both samples can be seen in Fig.~\ref{fig:X-ray binned flux} and ~\ref{fig:Appendix column density}, which show a removal of the more obscured sources within the bins and no CT sources remaining in the \citetalias{pfeifle2022bass} sample. It is evident that the luminosity cut works as an effective cut in $N_{\mathrm{H}}$, removing any CT sources from the sample. The flux cut only removed the very heavily obscured sources from the samples, with a similar distribution in $N_{\mathrm{H}}$ to the NSS80 sample. The X-ray flux cut does not significantly change the distribution of sources within each bin as evidenced by Fig.~\ref{fig:Appendix flux bin plot}.

The \citetalias{Asmus_diagnostic} sample was concluded to be statistically similar to the NSS80 in the column density distribution, even with the luminosity and flux cuts applied. The result implies that the \citetalias{Asmus_diagnostic} sample difference with the NSS80 is not solely due to the higher average X-ray luminosity within the sample. The most likely contributing factor is the discussed nuclear MIR luminosity, instead of observed, which reduces the influence of the host galaxy from a source.

\bsp	
\label{lastpage}
\end{document}